\documentclass[]{aastex631}

\begin{document}

\title{Fantastic Fits with \texttt{fantasy} of Active Galactic Nuclei Spectra \\
 - Exploring the Fe II emission near the H$\alpha$ line}

\correspondingauthor{Dragana Ili\'c}
\email{dilic@matf.bg.ac.rs}

\author[0000-0002-1134-4015]{Dragana Ili\'c}
\affiliation{Department of Astronomy, University of Belgrade - Faculty of Mathematics, Studentski trg 16, 11000 Belgrade, Serbia}
\affiliation{Humboldt Research Fellow, Hamburger Sternwarte, Universit{\"a}t
Hamburg, Gojenbergsweg 112, 21029 Hamburg, Germany}

\author[0000-0002-2514-0793]{Nemanja Raki\'c}
\affiliation{Physics Department, Faculty of Natural Sciences and Mathematics, University of Banjaluka,\\ Mladena Stojanovi\'ca 2, 78000 Banjaluka, RS, Bosnia and Herzegovina}

\author[0000-0003-2398-7664]{Luka \v C. Popovi\'c}
\affiliation{Astronomical Observatory, Volgina 7, 11000 Belgrade, Serbia}
\affiliation{Department of Astronomy, University of Belgrade - Faculty of Mathematics, Studentski trg 16, 11000 Belgrade, Serbia}

\begin{abstract}

In this study, a refined approach for multicomponent fitting of active galactic nuclei (AGN) spectra is presented utilizing the newly developed \texttt{Python} code \texttt{fantasy} (fully automated python tool for AGN spectra analysis). AGN spectra are modeled by simultaneously considering the underlying broken power-law continuum, predefined emission line lists, and an Fe II model, which is here extended to cover the wavelength range 3700 - 11000 \AA. The Fe II model, founded solely on atomic data, effectively describes the extensive emission of the complex iron ion in the vicinity of the H$\gamma$ and H$\beta$ lines, as well as near the H$\alpha$ line, which was previously rarely studied.
The proposed spectral fitting approach is tested on a sample of high-quality AGN spectra from the Sloan Digital Sky Survey (SDSS) Data Release 17.
The results indicate that when Fe II emission is present near H$\beta$, it is also detected redward from H$\alpha$, potentially contaminating the broad H$\alpha$ line wings and thus affecting the measurements of its flux and width. The production of Fe II emission is found to be strongly correlated with Eddington luminosity and appears to be controlled by the similar mechanism as the hydrogen Balmer lines. The study highlights the benefits of fitting AGN type 1 spectra with the \texttt{fantasy} code, pointing that it may be used as a robust tool for analyzing a large number of AGN spectra in the coming spectral surveys.

\end{abstract}


\section{Introduction} \label{sec:intro}

Active galactic nuclei (AGN) spectra can be very complex, with underlying emission from the stellar component of the host galaxy, continuum emission predominantly from the accretion disk, and strong broad and narrow emission lines originating from regions at a wide range of distances from the central supermassive black hole \citep[SMBH, see e.g.][]{2013peag.book.....N}. Disentangling the complex optical spectra of AGN is an important part of AGN research to understand the physical processes behind continuum and emission line production \citep[][]{2017A&ARv..25....2P}. In addition, robust and reliable extraction of spectral parameters in type 1 AGN\footnote{Type 1 AGN, also known as unobscured AGN, show broad emission lines in their UV, optical and NIR spectra.}, such as the width and flux of the broad emission lines or the underlying continuum flux, is of importance for their application in estimating physical parameters such as the mass of the SMBH or the Eddington ratio \citep[][]{2011ApJS..194...45S, 2019ApJS..243...21L}. Both are needed to understand AGN and their role in galaxy evolution \citep[e.g.][]{2013ARA&A..51..511K}. The two best known broad emission lines are the H$\beta$ and H$\alpha$ lines, which are well studied and used for {a} large number of type 1 AGN \citep[e.g.,][]{Greene05, 2011ApJ...739...28X, 2011ApJS..194...45S, 2019ApJS..243...21L, 2020ApJS..249...17R}. 

We are still far from fully understanding the physical processes of the plasma in the broad line region (BLR), such as {what the densities and temperatures are} \citep[][]{2003ApJ...599..140P,2020Atoms...8...94M}, {and} how to predict the observed emission line ratios \citep[e.g.,][]{2012A&A...543A.142I,2020MNRAS.494.1611N}, {and} understand {the} production of diffuse BLR continuum emission \citep[][]{2019NatAs...3..251C}, {and} the presence and role of dust \citep[][]{2016ApJ...832....8B,2022arXiv221111022C}, or determine the location and origin of Fe II emission \citep[e.g.,][]{2004ApJ...615..610B,2022AN....34310112G}. For sure, the promising approach for these investigations is to exploit large spectral data sets and provide catalogues of their spectral properties, {as} has been done for more than half a million of quasars from the Sloan Digital Sky Survey \citep[SDSS,][]{2020ApJS..249...17R}.

Some of the challenges in extracting pure broad H$\beta$ and H$\alpha$ line profiles and measuring their spectral parameters, such as the full width at half maximum (FWHM) and line fluxes, lie in subtracting the contribution of the host galaxy stellar emission, estimating the underlying AGN continuum emission, and identifying and subtracting narrow and other satellite lines. This can be particularly difficult for type 1 AGNs with strong Fe II emitters, such as the narrow-line Seyfert 1 \citep[NLSy1, see e.g.,][]{2017ApJS..229...39R,2022MNRAS.516.2374P}. Contamination by Fe II is probably the most difficult to deal with. It is known that the complex Fe II ion generates thousands of line transitions \citep[][]{1985ApJ...288...94W, 1998ApJ...499L.139S, 2003ApJS..145...15S,2021ApJ...907...12S}. Therefore, these lines are typically blended and difficult to identify in AGN spectra, so there are several Fe II templates that can be used by the community \citep[for recent review on different Fe II {templates}, see][and references therein]{2022ApJS..258...38P}. However, most templates do not focus on the region near the H$\alpha$ line,
 which can be contaminated by iron emission, especially in the line wings \citep{2004A&A...417..515V}. 

Moreover, recent observations of the specific transient phenomena of stellar disruption in galactic nuclei, so-called tidal disruption events \citep[TDEs, for a review see][]{2021ARA&A..59...21G} show that some of these events are strong Fe II emitters \citep[][]{2023A&A...669A.140P}. As pointed out in \cite{2021ApJ...920...56F}, misleading identification  of emission line features can cause errors in the classification of these events. TDEs are becoming increasingly important because they provide a special opportunity to detect and study intermediate-mass black holes \citep[][]{2022NatAs.tmp..258G}.
With ongoing and upcoming large sky surveys that aim to explore the transient sky, such as Zwicky Transient Factory \citep[ZTF][]{2019PASP..131f8003B} or Vera C. Rubin Legacy Survey in Space and Time \citep[LSST][]{2019ApJ...873..111I}, 
the number of detected transients will increase rapidly, as will their spectroscopic-optical follow-up with either single 
campaigns or dedicated surveys, such as a very successful Public ESO Spectroscopic Survey for Transient Objects \citep[PESSTO][]{2015A&A...579A..40S}, the already accepted extragalactic community surveys\footnote{https://www.4most.eu/cms/science/extragalactic-community-surveys/} as a part of the 4-metre Multi-Object Spectrograph Telescope \citep[4MOST][]{2012SPIE.8446E..0TD}, or the forthcoming Manuakea Spectroscopic Explore \citep[MSE][]{2019arXiv190404907T}.

{Hence}, there is a need for software packages that can perform modelling, fitting, and analysis of AGN spectra in an automated manner. There are already a few publicly available codes, namely: i) Quasar Spectral Fitting package \citep[QSFIT ][]{2017MNRAS.472.4051C}, an IDL based code for fitting all AGN emitting components simultaneously; ii) \texttt{Python} QSO fitting code \citep[PyQSOFit][]{2018ascl.soft09008G, 2019MNRAS.482.3288G}, designed for fitting quasar spectrum and additionally performs Monto-Carlo iterations using flux randomization to estimate uncertainties; iii) Sculptor, an interactive graphical user interface written in \texttt{Python} for general astronomical spectral analysis, with a special extension for quasar spectra \citep[][]{2022ascl.soft02018S}.
Nevertheless, other open-source packages are needed that are tailored to model optical AGN spectra and are easy to use. 
Here we use the recently developed \texttt{Python} code \texttt{fantasy} \citep[Fully automated python tool for AGN spectra analysis][]{ilic20,rakic22}, which is an updated approach to multicomponent fitting of AGN spectra. The main advantages of the code are: i) the ability to fit a wide range of wavelengths simultaneously (e.g., from H$\delta$ to H$\alpha$), ii) the selection of lines from predefined line lists with the option to easily insert user-defined line lists, and iii) the flexibility to model a wide variety of type 1 AGN spectra. A special feature of the code is the use of the extended model of iron emission in the wavelength range 3700 -- 11000 \AA \, for which initial concepts were presented in \cite{pop04} and further developed in \cite{kov10}.

Recently, during a transient event in an NLSy1 galaxy, the wavelength region redward from the H$\alpha$ line was heavily contaminated by iron emission \citep[for details see][]{2023A&A...669A.140P}. Most available iron templates focus on the H$\beta$ region, since this line is most commonly observed in distant AGN and is widely used for SMBH mass estimates from single-epoch observations \citep[for a review see][]{pop20}. With new instruments focusing on the NIR spectrum, such as the James Web Space Telescope, observation of the H$\alpha$ line in distant quasars is becoming possible, and the need for more Fe II templates and models covering this wavelength band is evident. This has motivated us to investigate iron emission, which is known to exist in the vicinity of the H$\alpha$ line but has not yet been studied in detail except in \cite{2004A&A...417..515V}. In addition, we will demonstrate the importance of simultaneously fitting the spectra of type 1 AGNs in a broader wavelength range, including emission line components along with the underlying continuum and Fe II emission model. We will investigate the physical properties of regions emitting broad lines, with particular attention to the  Fe II emission near the H$\alpha$ line.
For this purpose, we use a selection of optical spectra of type 1 AGN obtained from the public database of the SDSS latest Data Release 17\footnote{ \url{https://www.sdss4.org}}, as well as as the publicly available optical spectra of I Zw 1, a well-known NLSy1, typically used to demonstrate the suitability of iron templates in AGN.

The paper is organized as follows. In Sect. 2, we describe the data set used. In Sect. 3 we present our extended semi-empirical model of Fe II emission and describe the main functionalities of \texttt{fantasy} code, whereas in Sect. 4 we present the results and provide relevant discussion. Finally, in Sect. 6 we list our conclusions. We assume a cosmology with H$_0=67$ km s$^{-1}$ Mpc$^{-1}$, $\Omega_m = 0.32$ and $\Omega_\Lambda = 0.68$ to calculate the luminosity distance to studied objects.

\section{Data sample}

We select a sample of type 1 AGN from the latest SDSS Data Release 17 \citep[DR17,][]{dr17}, where we specify as selection criteria the high signal-to-noise ratio of the optical spectra 
(S/N $>$ 35)  and the redshift $z<0.4$. The first criterion excludes the presence of noisy spectra that may significantly bias the modelling of complex AGN spectra, especially iron emission. The second criterion ensures that both H$\beta$ and H$\alpha$ are included. The DR17 is a fourth data release from the fourth phase of the survey (SDSS-IV), which contains the complete data set of optical single-fibre spectroscopy of the SDSS\footnote{\url{www.sdss.org/dr17/spectro/}} through January 2021 \citep[][]{2013AJ....146...32S}. The selection process and the characteristics of the selected sample are the same as in \citet{rakic22}, who used DR16 data \citep{dr16}. The query yielded 676 objects,  of which 21 were discarded either because H$\alpha$ emission was absent or distorted, or {objects} were misclassified as type 1 AGN {while} they are rather type 2 as no broad component is seen. The remaining 655 SDSS AGN type 1 objects were further investigated (referred to as the SDSS sample).

One way to represent the distribution of type 1 AGN is to use a quasar main sequence diagram, defined with FWHM H$\beta$ - $R_{\rm Fe II}$\footnote{$R_{\rm Fe II}$ is a measure of the Fe II optical emission, defined as the ratio of the equivalent widths of Fe II emission in the wavelength range 4435–4685 \AA \, and H$\beta$ broad line.} space \citep[][]{sulentic00,2014Natur.513..210S,2016ApJ...818L..14D,marziani22}. These studies {have} shown that $R_{\rm Fe II}$ emission is a proxy for accretion strength in AGN, roughly also indicated by the Eddington ratio \citep[see also][]{2011ApJ...736...86D}. The quasar main sequence is mainly occupied by two large groups of objects, populations A and B, defined according to the full width at half maximum (FWHM) of H$\beta$ line \citep[e.g.,][]{sulentic00,2002ApJ...566L..71S,marziani22}. These two populations have different physical properties and possibly different orientation \citep[][]{marziani18,marziani22}. Population B objects have broader emission lines (FWHM $\geq$ 4000 km s$^{-1}$), higher inclination angle, low Eddington ratio, and weak Fe II emission ($R_{\rm Fe II} \leq 0.5$), whereas population A, occupying the other side of the diagram, have narrower broad emission lines, lower inclination, higher Eddington ratio, and strong Fe II \citep[see a recent review][]{marziani22}. The extreme end of the objects of population A, which show very strong iron emission  (R$_{\rm FeII} >$ 1), are also of great interest  \citep[][]{marziani22}.

\begin{figure}
\centering
\includegraphics[width=0.5\columnwidth]{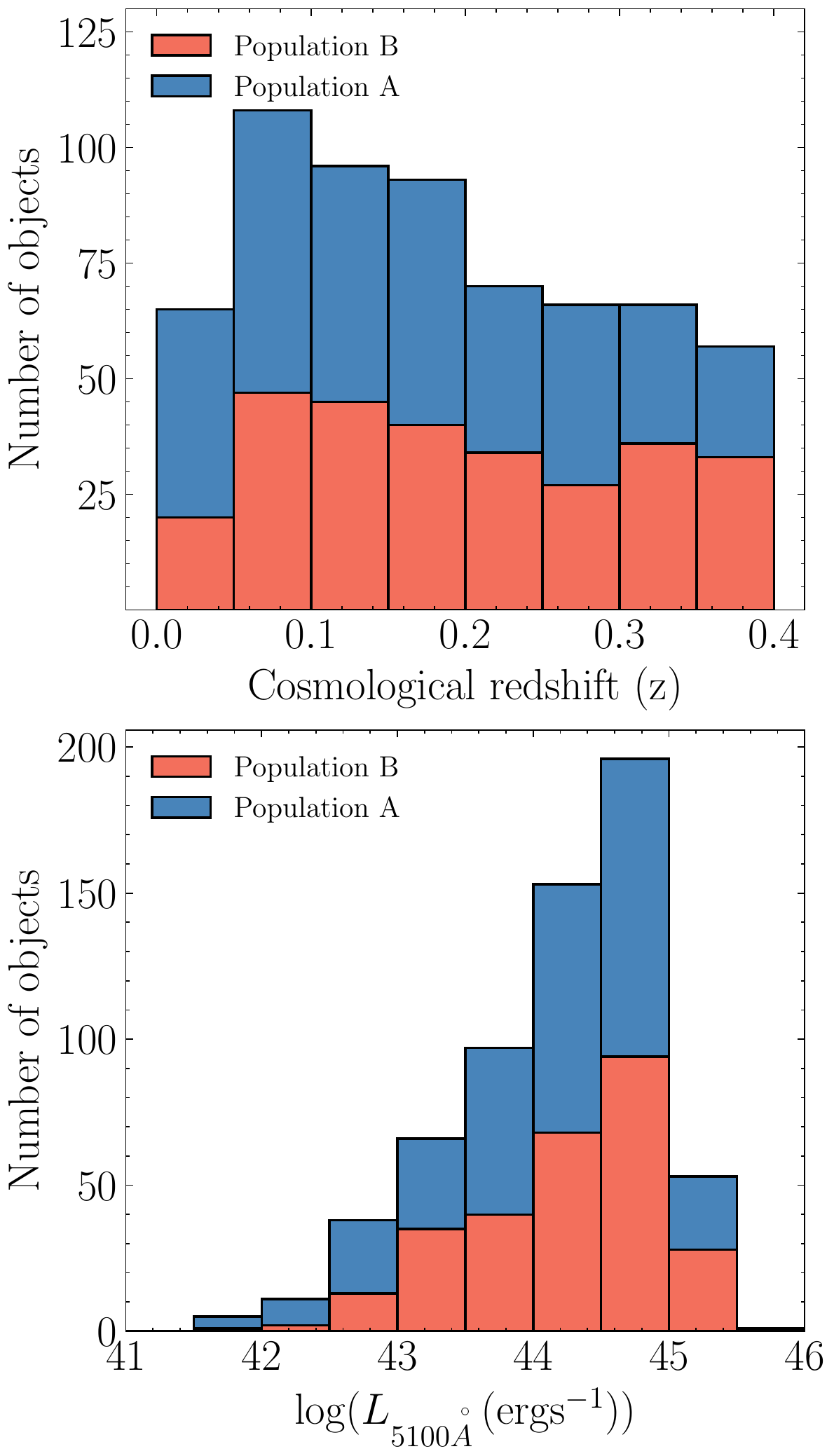}
\caption{Distribution of SDSS sample of 655 objects with the cosmological redshift (upper panel) and continuum luminosity at 5100 \AA\, (bottom panel) presented with stacked bars of pop A (red) and pob B (blue) sub-samples.
\label{fig:hist}}
\end{figure}

Following the above, our SDSS sample of 655 objects was divided into 362 objects from population A (referred to as the ``pop A'' sample in the remainder of this text) and 293 objects from population B (``pop B'' sample). A subset of 105 objects from the pop A sample belongs to the so-called extreme population A (R$_{\rm FeII} >$ 1). This sample is treated separately and we refer to it as the ``xA'' sample.
Fig. \ref{fig:hist} presents the distributions of the selected SDSS sample with the cosmological redshift $z$ (upper panel) and continuum luminosity at 5100 \AA\, (lower panel), shown as stacked bars of pop A (red) and pop B (green) sub-samples. All samples (total, pop A, pop B) are almost uniformly distributed across the redshift, showing {a} slight decrease towards higher redshift \citep[seen also in][who studied a sample of type 1 AGN with redshift $<0.35$]{2019ApJS..243...21L}. The continuum luminosity distribution is asymmetric toward higher-luminosity, which is typical for type 1 AGN from SDSS \citep[][]{2019ApJS..243...21L}, with most objects (95\%) in the range log($L_{5100}$) = [42.5-45.3] erg s$^{-1}$, with the median of 44.3.

To demonstrate the importance of simultaneous multi-component fitting of AGN spectra, in particular to extract Fe II emission, we also test spectral fitting on a publicly available optical spectrum of the well-known NLSy1 I Zw1 \citep[taken from][]{2006ApJ...650...57T}\footnote{A spectrum was retrieved via the NASA/IPAC Extragalactic Database (NED), \url{https://ned.ipac.caltech.edu}, see also \url{http://www.ioa.s.u-tokyo.ac.jp/~kkawara/quasars/}.}, which is widely used to construct and test templates and models of Fe II emission in AGN.

\begin{figure}
\centering
\includegraphics[width=0.9\columnwidth]{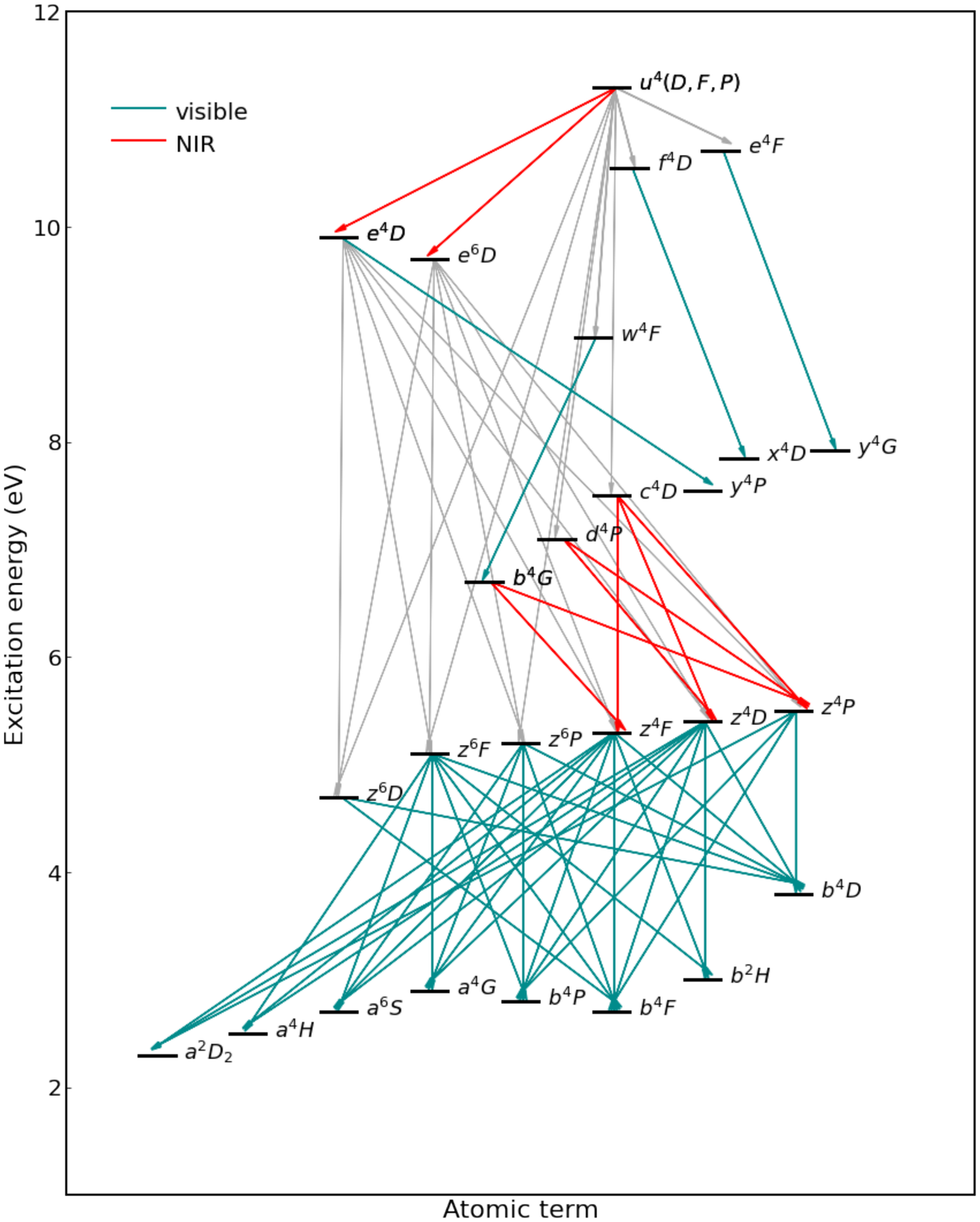}
\caption{Partial Grotrian diagram for Fe II, showing the transitions considered in the presented model of Fe II emission in the wavelength range 3700-11000 \AA. The upper levels u$^4$(D,F,P) are populated by Ly$\alpha$ photons, from which the cascades to lower levels are shown: red lines show the NIR transitions (bumps at 9200 \AA\, and 1$\mu$m) and dark cyan lines represent the transitions in the optical band (centered at 4570 \AA\, and 5270\AA\, which are the two strongest bumps around H$\beta$). The gray lines represent the transitions responsible for populating the upper levels, mainly through UV emission, to illustrate the path. 
\label{fig:grotrian}}
\end{figure}

\begin{figure*}
\centering
\includegraphics[width=\textwidth]{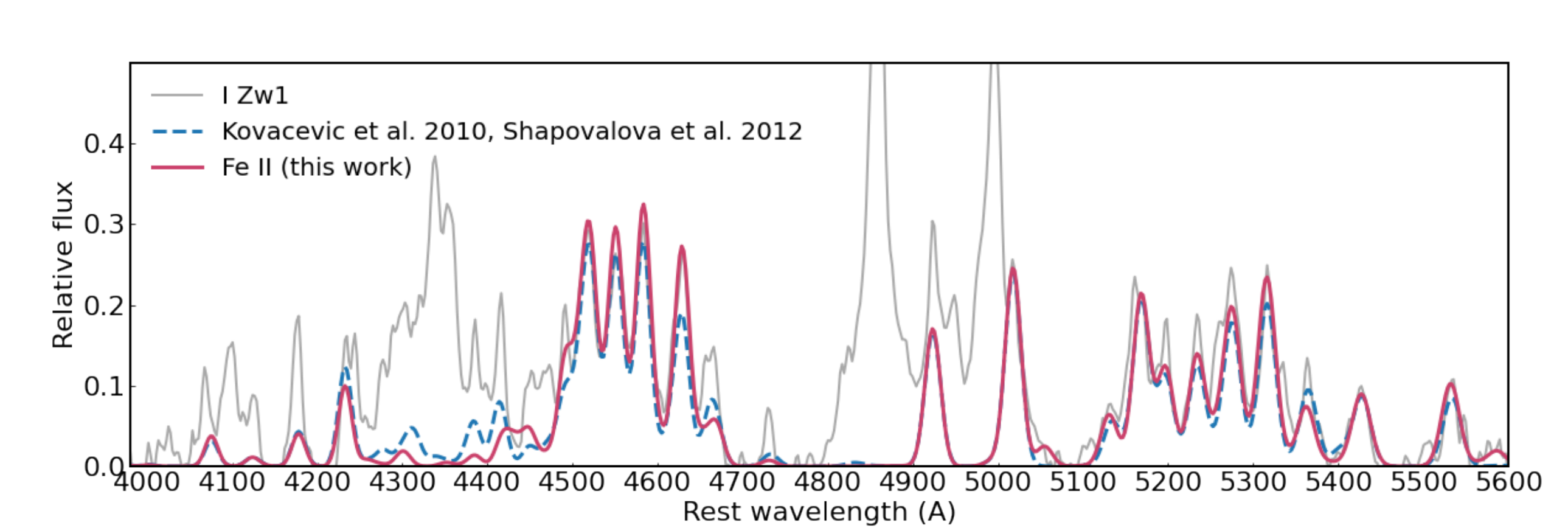}
\caption{Comparison of the semi-empirical Fe II model presented in \cite{kov10} and \cite{2012ApJS..202...10S} with the model of Fe II emission presented in this work, which relies solely on the atomic data.
\label{fig:temp_kov}}
\end{figure*}

\begin{figure*}
\centering
\includegraphics[width=\textwidth]{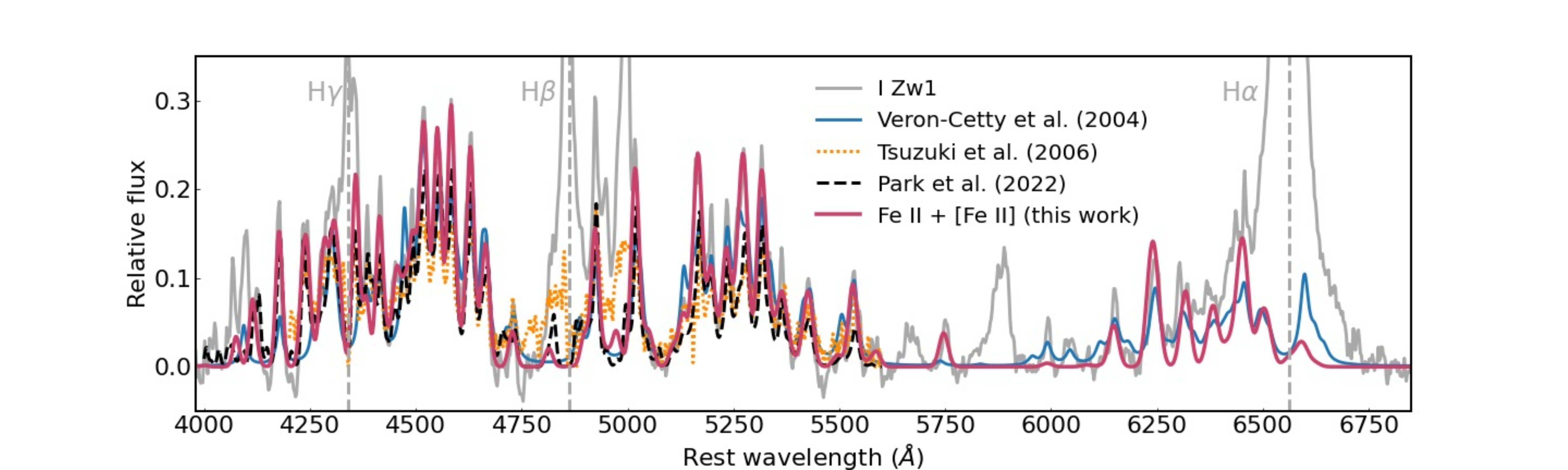}
\caption{Different Fe II templates compared to the I Zw 1 spectrum from which the continuum emission was subtracted. The Fe II model presented here is the result of fitting the observations and also includes [Fe II] lines that are strong in this object (see Section 3.2.1 for details).
\label{fig:temp_lit}}
\end{figure*}

\section{Methods and analysis}

In this section we present the approach to modeling the complex spectra of type 1 AGN using the code \texttt{fantasy} and the description of the extended Fe II emission model.

\subsection{Extension to the atomic model of Fe II emission} \label{sec:feii}

For interpreting the emission within the type 1 AGN spectra, one important requirement is to carefully reconstruct the Fe II emission. In addition, understanding the origin of the strong emission of a complex ion of one-time ionized iron in AGN is a continuing quest \citep[e.g.,][]{1980ApJ...236..406N,1987MNRAS.229P...1P,1999ApJS..120..101V,2004ApJ...611...81S,2019ApJ...882...79P,2021ApJ...907...12S,2022AN....34310112G}. Fe II emission is present in a broad spectral range and typically contaminates other broad lines from the UV \citep[e.g., Mg II line][]{2001ApJS..134....1V, pop19}, through optical \citep[e.g., H$\beta$ and H$\alpha$ lines][]{2004A&A...417..515V, 2022ApJS..258...38P} to near-infrared bands \citep[e.g., Pa$\gamma$ line][]{2000ApJ...539..166R,2008ApJS..174..282L, 2012ApJ...751....7G, 2016ApJ...820..116M}. Their investigations are important for understanding the physics of the broad line region.

Usually, iron emission is modeled with empirical templates \citep{1992ApJS...80..109B,  2001ApJS..134....1V, 2004A&A...417..515V, 2006ApJ...650...57T, 2022ApJS..258...38P}. These templates consider that the relative intensities of identified Fe II lines are proportional to empirical ratios measured in AGN spectra whose broad lines are narrow enough, so that the Fe II features are more easily separated from other emission lines, i.e. NLSy1. Some Fe II templates are governed by the identification of emission lines within the same important multiplets \citep{2016ApJ...820..116M, 2020MNRAS.494.4187M}. The semi-empirical approach to modelling the Fe II emission was presented in \cite{kov10} and updated in \cite{2012ApJS..202...10S}. Their semi-empirical model covers the wavelength range  4000 - 5500 \AA, and is mainly based on the atomic parameters of the strongest iron transitions. In short, the model consists of Fe II line-sets grouped according to the same lower energy level in the transition with line intensities connected by the line transition oscillatory strengths, for a given excitation temperature, usually assumed to be 10$^4$K. For few line groups, the relative line intensities were measured from I Zw 1 \citep{kov10}, which made this model a semi-empirical one. For a {detailed} presentation of most relevant  Fe II templates, their properties, applications, and comparisons, we refer to the recent review presented in \cite{2022ApJS..258...38P}. It is worth noting that given the complex energy level structure of the Fe II ion and a huge number of transitions producing rich spectrum in the UV, optical and NIR regions, it remains challenging to use in practice fully theoretical Fe II templates \citep[such as][]{2003ApJS..145...15S, 2008ApJ...675...83B}. We emphasize again that most Fe II templates focus on the H$\beta$ region, which is the most studied broad emission line in AGN.

\begin{figure*}
\centering
\includegraphics[width=1\textwidth]{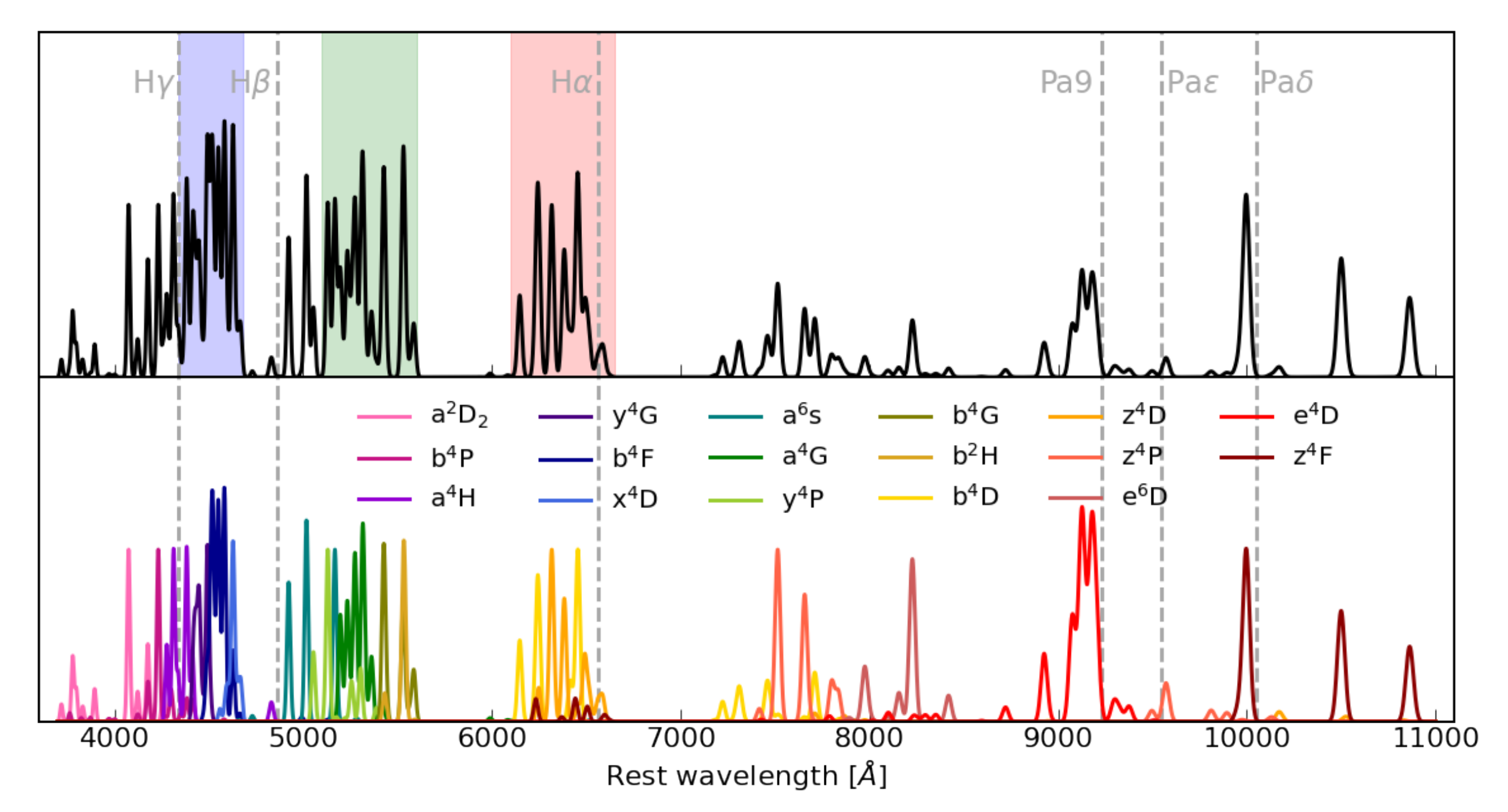}
\caption{Fe II model in the wavelength range 3700 - 11000 \AA, with atomic groups indicated in the bottom panel. The position of the hydrogen Balmer and Paschen series are also indicated with the dashed vertical lines, whereas the shaded areas indicate the position of iron bands used later in the analysis: Fe II blue (4340--4680 \AA), Fe II green (5100--5600 \AA), and Fe II red (6100--6650 \AA).
\label{fig:temp_dif}}
\end{figure*}

\begin{deluxetable}{ccccccc}
\tabletypesize{\scriptsize}
\tablewidth{0pt} 
\tablecaption{Example of Fe\,{\footnotesize II} transitions used in
    the model for the lower level $a^6S$ (energy 2.891 eV) group with
    the reference line set to a relative intensity of 1.000. \label{tab:Fe}}
\tablehead{
\colhead{{Wavelength (air)}} & \colhead{log(gf)}& \colhead{transition} & \colhead{E(low)} & \colhead{E(up)} & \colhead{Relative}& \colhead{Ref.} \\
\colhead{\AA} & \colhead{} & \colhead{lower -- upper} & \colhead{eV} & \colhead{eV} & \colhead{intensity} &
} 
\startdata 
4593.83	&	-4.923	&	$	a^6S_{5/2}	-	z^4F_{5/2}	$	&	2.891	&	5.589	&	0.000  &	(3)	\\
4601.38	&	-4.428	&	$	a^6S_{5/2}	-	z^4D_{3/2}	$	&	2.891	&	5.585	&	0.001  &	(3)	\\
4656.98	&	-3.630	&	$	a^6S_{5/2}	-	z^4D_{5/2}	$	&	2.891	&	5.553	&	0.007  &	(3)	\\
4663.71	&	-3.820	&	$	a^6S_{5/2}	-	z^4F_{7/2}	$	&	2.891	&	5.549	&	0.004  &	(1)	\\
4731.45	&	-2.921	&	$	a^6S_{5/2}	-	z^4D_{7/2}	$	&	2.891	&	5.511	&	0.035  &	(2)	\\
4923.93	&	-1.561	&	$	a^6S_{5/2}	-	z^6P_{3/2}	$	&	2.891	&	5.408	&	0.810  &	(2)	\\
5018.44	&	-1.400	&	$	a^6S_{5/2}	-	z^6P_{5/2}	$	&	2.891	&	5.361	&	1.170  &	(2)	\\
5169.03	&	-1.466	&	$	a^6S_{5/2}	-	z^6P_{7/2}	$	&	2.891	&	5.289	&	1.000  &	(2)	\\
5256.94	&	-4.250	&	$	a^6S_{5/2}	-	z^6F_{5/2}	$	&	2.891	&	5.249	&	0.002  &	(3)	\\
5284.11	&	-3.121	&	$	a^6S_{5/2}	-	z^6F_{7/2}	$	&	2.891	&	5.237	&	0.022  &	(2)	\\
6369.46	&	-4.253	&	$	a^6S_{5/2}	-	z^6D_{3/2}	$	&	2.891	&	4.837	&	0.001  &	(3)	\\
6432.68	&	-3.708	&	$	a^6S_{5/2}	-	z^6D_{5/2}	$	&	2.891	&	4.818	&	0.005  &	(3)	\\
6516.08	&	-3.450	&	$	a^6S_{5/2}	-	z^6D_{7/2}	$	&	2.891	&	4.793	&	0.009  &	(3)	\\
\enddata
\tablecomments{Columns give line transition wavelength in air (in
    \AA), oscillatory strength, the configuration of lower and upper
    energy level, corresponding energies (in eV),  and a relative
    intensity with respect to the reference line, {where}
    intensity is set to 1.000. References for the used oscillatory
    strength: 1 - \citep{1981JPCRD..10..305F}, 2 - \citep{kov10}, 3 -
    atomic database
    (\url{https://lweb.cfa.harvard.edu/amp/ampdata/kurucz23/sekur.
    html}). Table \ref{tab:Fe} is published in its
    entirety in the machine-readable format. A portion is shown here
    for guidance regarding its form and content.}

\end{deluxetable}

\begin{figure*}
\centering
\includegraphics[width=1\textwidth]{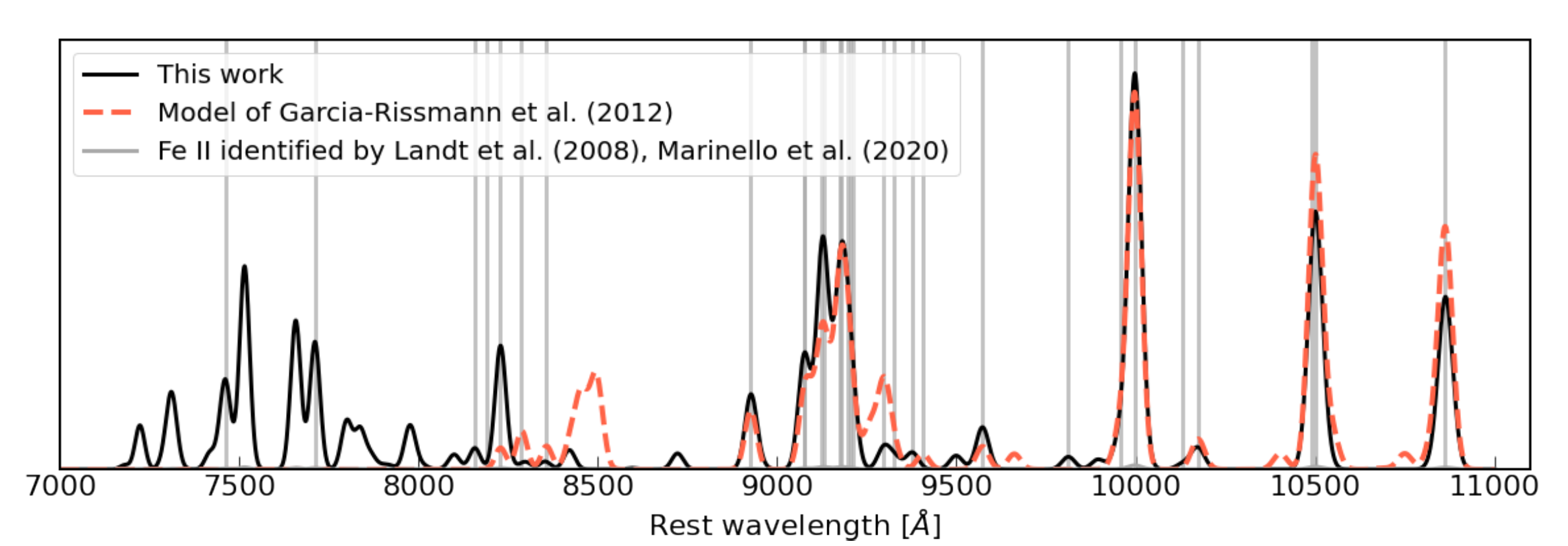}
\caption{Comparison of the Fe II model (black solid line) in the wavelength range 7000 - 11000 \AA, with the Fe II model provided for the wavelength range 8200 - 11400 \AA\, in \cite{2012ApJ...751....7G} (red dashed line, see text for details).  The position of Fe II line identified in \cite{2008ApJS..174..282L,2020MNRAS.494.4187M} are marked with vertical lines.
\label{fig:temp_NIR}}
\end{figure*}

In this work, our main motivation was to develop the Fe II model that is extended to include the H$\alpha$ line region, which is also strongly populated by Fe II lines in NLSy1 objects \citep[e.g.,][]{2006ApJ...650...57T, 2011ApJ...736...86D, 2022ApJS..258...38P}, and could be important in transient event, {such} as TDEs \citep[see][and reference therein]{2023A&A...669A.140P}. Therefore, based on the same approach and assumptions presented in \cite{kov10}, here we construct the Fe II model in the 3700 - 11000 \AA \, wavelength range. In addition, here we revise the group of so-called high-excitation lines, for which the line ratios have been measured from I Zw 1 spectrum in \cite{kov10}. It follows that the proposed model of Fe II is based solely on atomic data and the assumed excitation temperature. The individual line profiles in this model of Fe II are assumed to be Gaussians with the same width and shift. The lines are grouped to have the same lower level of transition, and all lines in a single group have a fixed intensity (calculated from the atomic data) relative to the strongest line in the atomic group. This leaves only the intensities of the strongest lines as free parameters. It has been long known that various Fe II optical multiplets {do not have} the same relative intensities in different objects \citep[see e.g.,][]{1993A&A...272...25V, 2004A&A...417..515V,2022ApJS..258...38P}. This multi-component model of Fe II emission has been shown to provide more flexibility for precise and careful fitting of diverse AGN spectra than monolithic empirical templates with a single overall intensity \citep[see e.g., the observations and analysis in][]{kov10,2012ApJS..202...10S, kov15,pop19}, where only the line width and shift are varied. In the following, we describe how the most important atomic transitions were selected for each atomic group.

We consider the possible path of electron transitions within the Fe II ion. In addition to the collisional excitation as one of the important {mechanisms} for the production of strong optical Fe II emission \citep[][]{2012ApJ...751....7G,2016ApJ...820..116M}, it has been discussed that the Ly$\alpha$ fluorescence is another relevant process for populating upper levels of Fe II \citep[][]{1987MNRAS.229P...1P,1998ApJ...499L.139S,2003ApJS..145...15S,2021ApJ...907...12S},
and thus responsible for NIR Fe II emission, and later contributing to the optical Fe II to a level of at least 20\% \citep[as shown in][]{2012ApJ...751....7G,2016ApJ...820..116M}. We start from the upper levels u$^4$(D,F,P), which  may be populated with Ly$\alpha$ photons (Figure \ref{fig:grotrian}). {From these, photons cascade down through transitions in the NIR (see Table \ref{tab:Fe})}, as also discussed and illustrated in \cite{2003ApJS..145...15S,2016ApJ...820..116M, 2020MNRAS.494.4187M}. These two groups of upper levels are responsible for populating the upper levels in the energy range of 4--7 eV through UV transitions (the Fe II emission in UV will be studied elsewhere). These are the known energy levels previously identified by \cite{kov10,2012ApJS..202...10S,2004A&A...417..515V} which give rise to two main optical Fe II bumps around H$\beta$ line centered at 4570 \AA\, and 5270 \AA\, (indicated with dark cyan lines in Fig \ref{fig:grotrian}).  We group the lines based on the same lower level of the transition, so that the line intensities are constrained  only by the
transition oscillatory strength $f$, which are listed $gf$ in Table \ref{tab:Fe}, where $g$ is the level statistical weight. As an example, 
{Table \ref{tab:Fe} {lists} the transitions identified within the a$^6$S group, and the complete list with all transitions {is} available in a machine-readable format in the online Journal.}

In selecting transition groups in the wavelength range near the H$\alpha$ line and beyond, we were guided by the Fe II transitions previously identified as strong, relying primarily on the work of \cite{2004A&A...417..515V,2022ApJS..258...38P} for the optical Fe II and \cite{2000ApJ...539..166R,2012ApJ...751....7G,2020MNRAS.494.4187M} for the NIR Fe II. Only the lines with  oscillatory strength (log$(gf) >-5$) were selected. The oscillatory strengths were adopted from \cite{kov10} and \cite{2012ApJS..202...10S}, whereas for the new transition groups they were taken from the atomic spectral line database from R. L. Kurucz\footnote{\url{https://lweb.cfa.harvard.edu/amp/ampdata/kurucz23/sekur.html}}. In {several} cases we have updated log($gf$) from the values given in \cite{1981JPCRD..10..305F} as they have been shown to give line ratios that better describe the observations. {The reference from which oscillatory strengths were taken is also given in Table \ref{tab:Fe}}. For the calculation of the line ratios, we used the excitation temperature of 10$^4$ K, which has been shown to represent well the region where these lines arise \citep{sulentic00,2012A&A...543A.142I}. The line ratios would not change significantly for small variations of excitation temperature \citep{kov10}.

We outline two important updates with respect to the semi-empirical Fe II model presented in \cite{kov10}, and further extended in \cite{2012ApJS..202...10S}:
\begin{enumerate}
    \item The wavelength range was extended to cover 3700 -- 11000 \AA, whereas originally it was focused on 4200 - 5600 \AA \, because this region around the H$\beta$ line was the most studied.
    \item Instead of using so-called high-excitation lines \citep[see discussion in][and their Table 3]{kov10},  whose ratios were previously measured from I Zw 1 spectrum, we have revised the list and assembled them into four atomic groups: y$^4$G, b$^4$G, x$^4$D, y$^4$P, as shown in Fig. \ref{fig:grotrian}. These groups are coming from the high-energy levels that are populated through {the} same paths as other levels (Fig. \ref{fig:grotrian}). The strongest lines in these groups were also identified by 
    \cite{kov10} (see their Table 3) and \cite{2004A&A...417..515V}.
\end{enumerate}

This makes it a full model of Fe II emission that relies only on the atomic data. It contains a total of 283 transitions, divided into 17 atomic groups, in the wavelength range of 3700 -- 11000 \AA\, (line transitions are listed in Table \ref{tab:Fe} and shown in Fig. \ref{fig:grotrian} and \ref{fig:temp_dif}, bottom panel).  In comparison, the previous semi-empirical model had 57 line transitions in the 4200 -- 5600 \AA, now the same range contains 125 transitions. However, the number of free parameters used in the fits has remained the same. Most of these lines are much weaker {(see Table \ref{tab:Fe})}, but these are all included since in iron-rich objects their contribution may not be negligible, and they do not burden the computation (see Section 3.2 for details on modeling). Nevertheless, the extended model of Fe II emission differs only slightly from the original semi-empirical of \cite{kov10} model as shown in Fig. \ref{fig:temp_kov}.

\begin{figure*}[ht!]
\centering
\includegraphics[width=\textwidth]{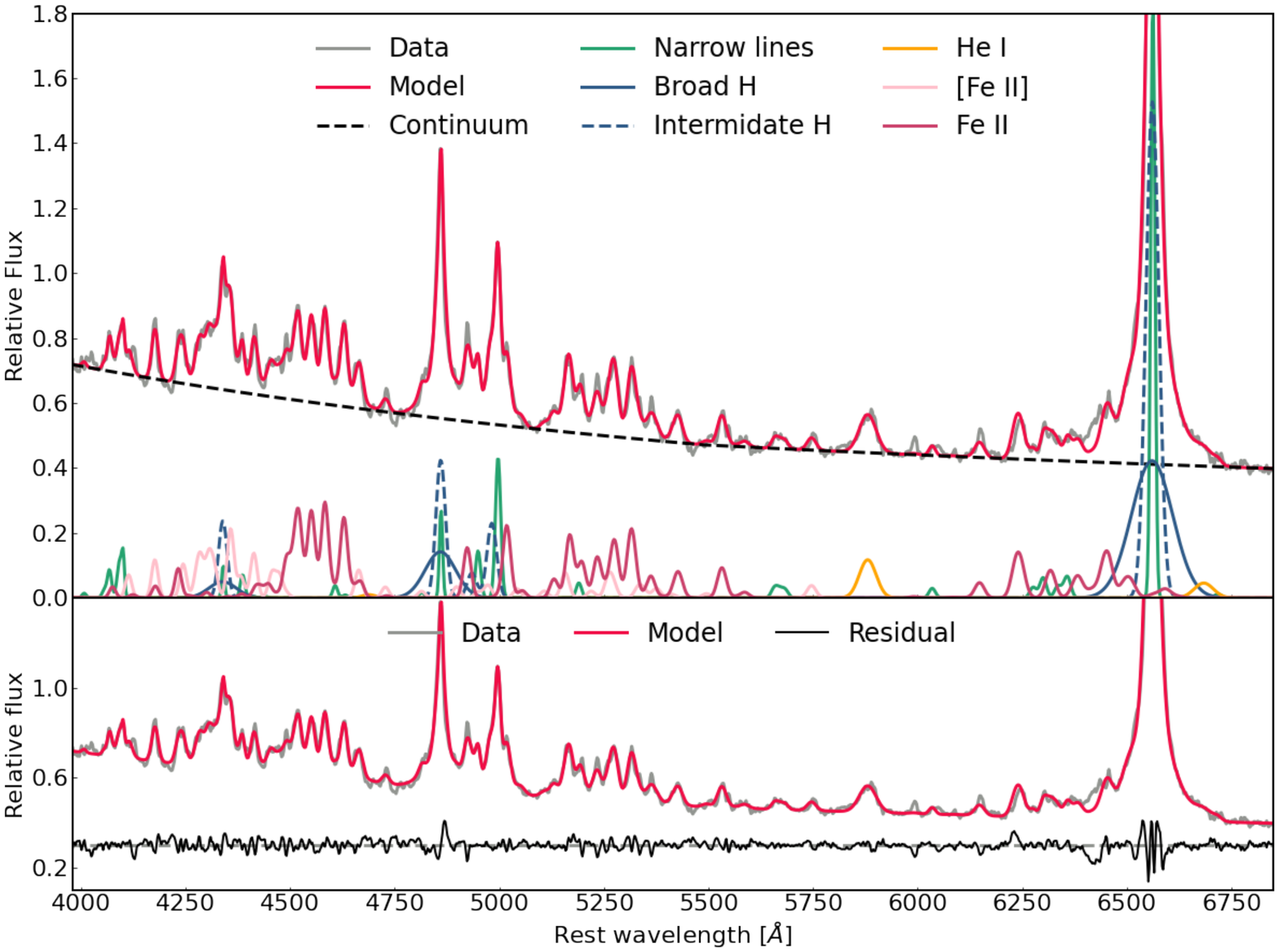}
\caption{Multi-component fitting with the  \texttt{fantasy} code of the I Zw 1 observed spectrum (gray line) in the $\lambda\lambda$4000-6800 \AA\, wavelength region. The model (red line) consists of: broken power-law (black dashed line), narrow lines (green solid line), broad (blue solid line) and intermediate (blue dashed line) components of Balmer lines: H$\alpha$, H$\beta$, H$\delta$, He I lines (yellow line), intermediate components of [O III] (blue dashed line), and Fe II model (dark-red line) and [Fe II] lines (light-red line). Bottom panel shows the zoomed-in observed (gray line), model (red line), and residual spectrum (black line).
\label{fig:fitZw}}
\end{figure*}

Figure \ref{fig:temp_lit} presents some widely used Fe II templates taken from the literature that are compared to the I Zw 1 spectrum, from which the continuum emission has been subtracted. The Fe II model presented in this paper is a result of the multi-component fitting (described in Section 3.2) and includes also [Fe II] lines which are strong in this and other similar iron-rich AGN. Full Fe II model in the wavelength range 3700 - 11000 \AA\, is presented in Fig. \ref{fig:temp_dif} with 17 atomic groups indicated on the bottom panel with different color. All lines are set to have a width of 1300 km s$^{-1}$, with intensities of reference lines arbitrary selected.

\begin{figure}
    \centering
    \includegraphics[width=0.5\columnwidth]{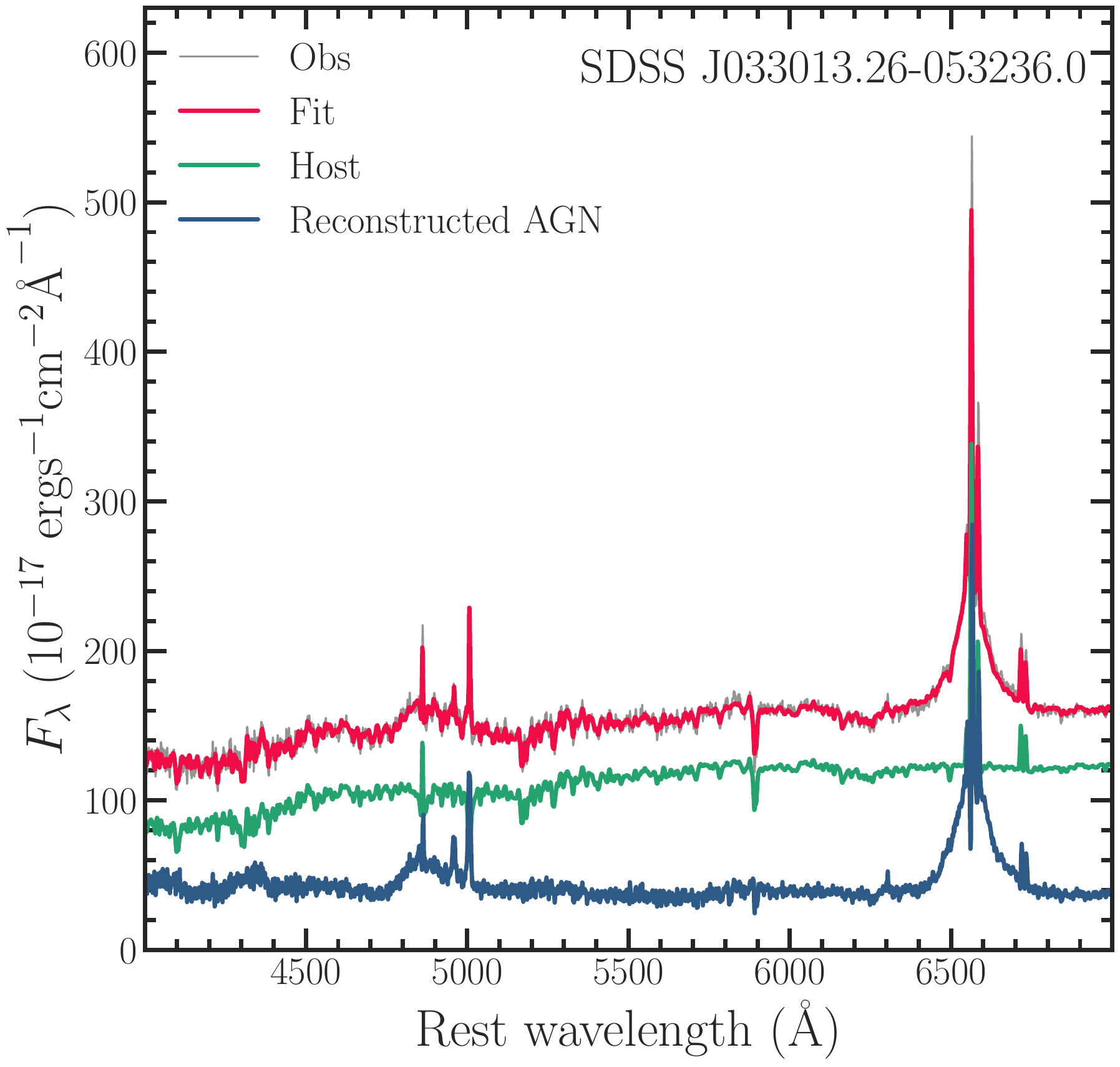}
    \caption{An example of the AGN component (blue line) reconstructed after subtracting the contribution of the stellar component of the host galaxy (green line) from an observed spectrum (gray line) of SDSS J033013.26-053236.0. The host-galaxy emission {contributes} on the level of 76\% to the observed continuum flux around 6000 \AA}. 
    \label{fig:host}
\end{figure}

In Figure \ref{fig:temp_NIR} we zoom in the NIR part of the Fe II model in the wavelength range 7000 - 11000 \AA. There {is} not much work dedicated to building the Fe II templates in NIR, mostly due to {the} observational limitation to obtain NIR spectra for distant AGN. In selecting the most dominant line transitions, we were governed by so-called well-known ``1$\mu$m'' Fe II features at $\lambda$9997, $\lambda$10501, and $\lambda$10863 \citep[][]{2000ApJ...539..166R}. We compare the proposed model with the Fe II model provided for the wavelength range 8200 - 11000 \AA\, by \cite{2012ApJ...751....7G} (Fig. \ref{fig:temp_NIR}, red dashed line), compiled from their best model for I Zw 1 (see their Tables 2 and 3).  Fe II line identified in \cite{2008ApJS..174..282L,2020MNRAS.494.4187M} are also marked with vertical lines. In this work, we focus on the spectra in 4000-7000\AA\, wavelength range due to {the} availability of SDSS spectra in this domain. 
The properties of 221 NIR iron emission in AGN spectra and further testing of the proposed model will be investigated in detail in a forthcoming publication.

\begin{figure}
    \centering
    \includegraphics[width=0.6\columnwidth]{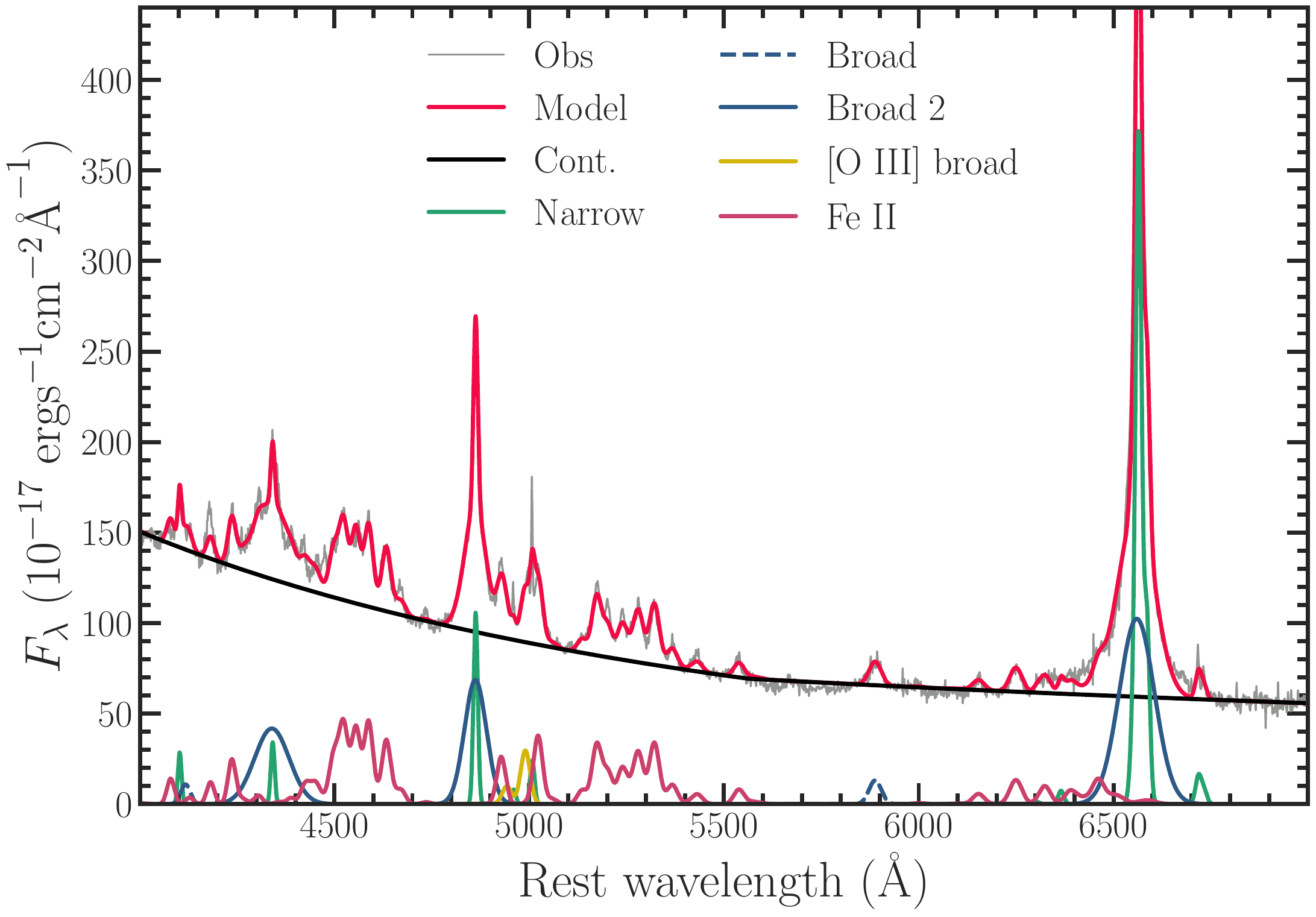}
    \includegraphics[width=0.6\columnwidth]{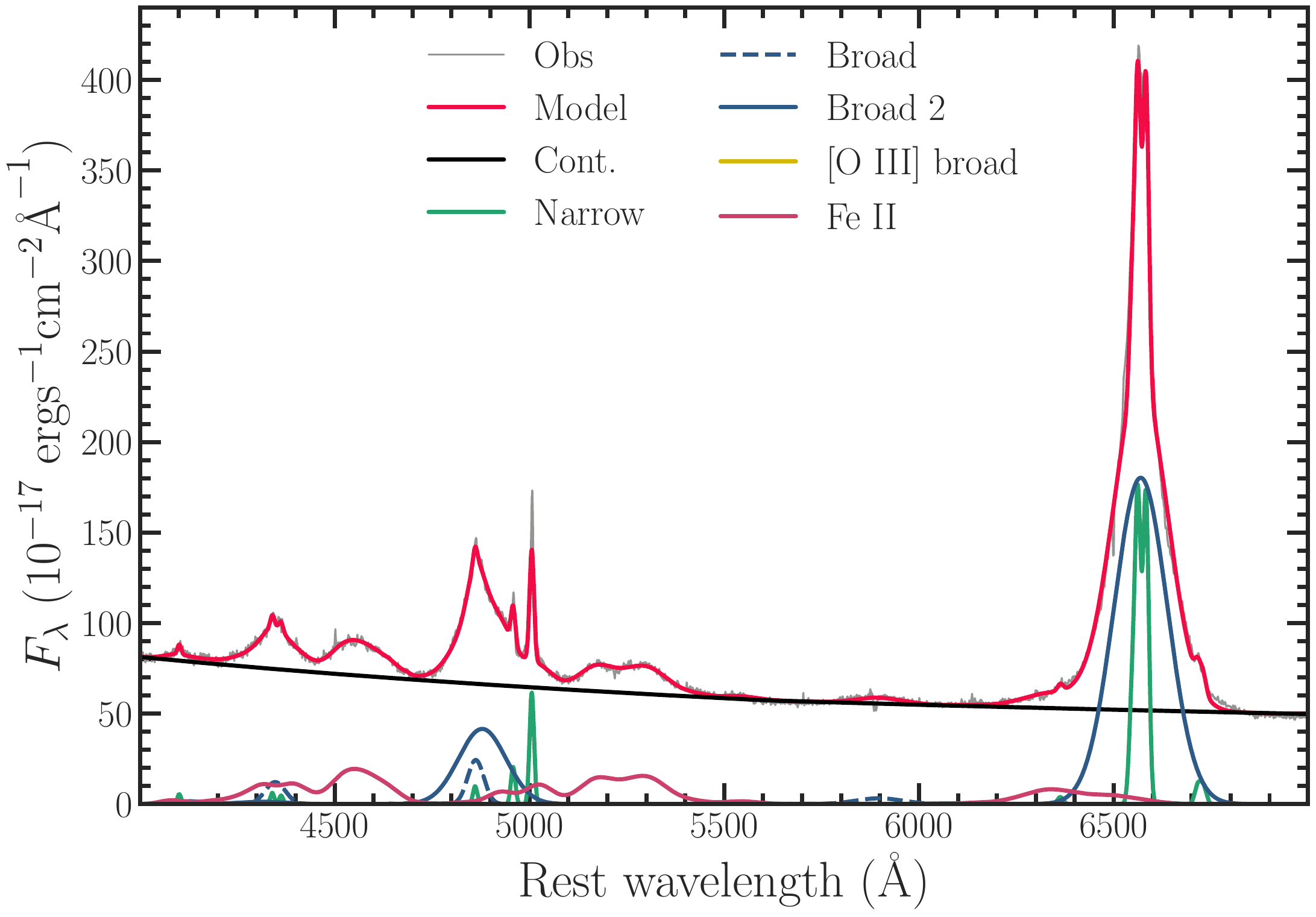}
    \includegraphics[width=0.6\columnwidth]{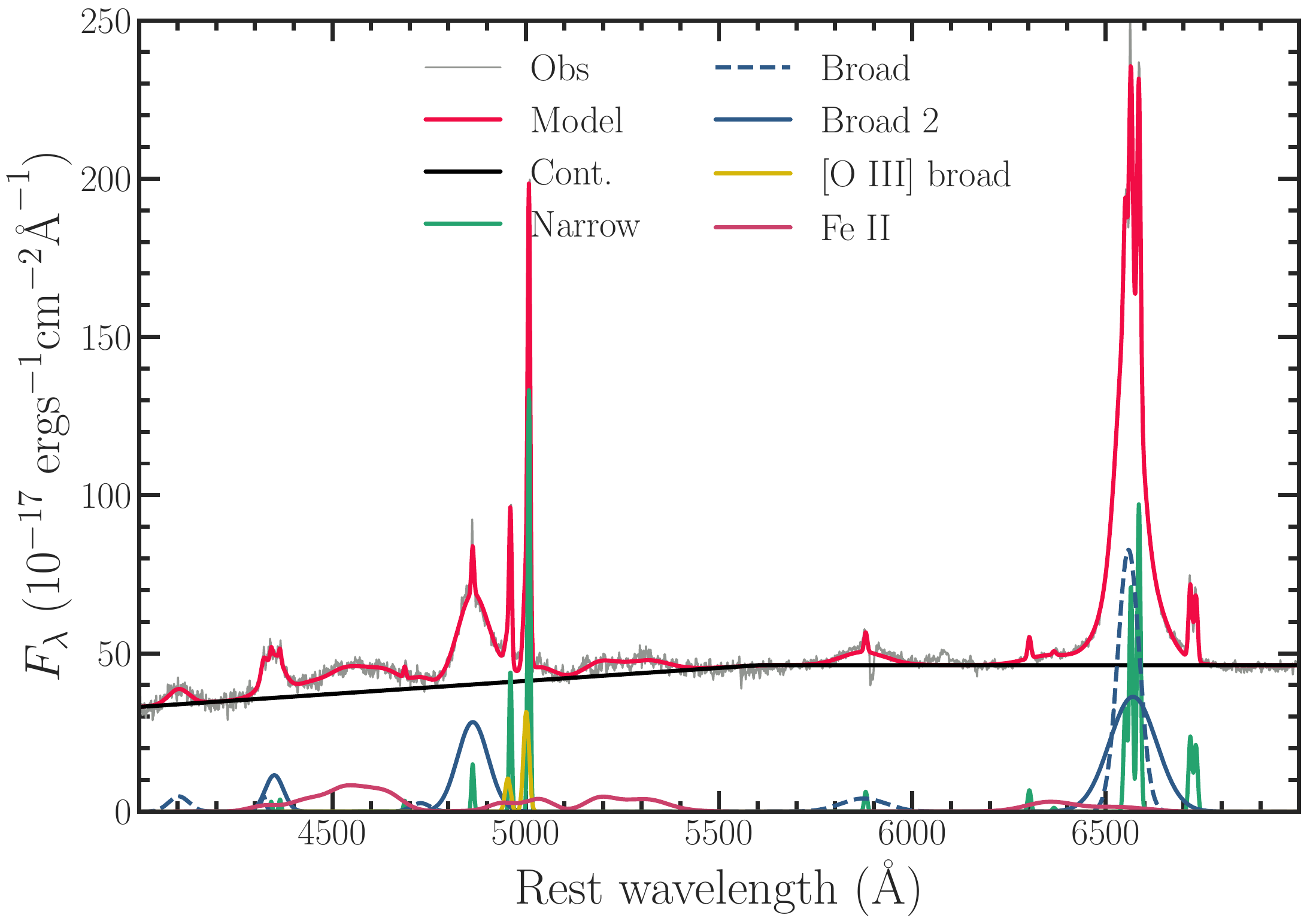}    
    \caption{The same as in Figure \ref{fig:fitZw} but for three cases of host-galaxy corrected spectra from SDSS sample with diverse spectral properties to illustrate the \texttt{fantasy} multi-component fittings: SDSS J010226.31-003904.5 (upper), SDSS J094620.86+334746.9 (middle) and SDSS SDSS J093641.05+101415.9 (bottom). All components used in the fitting model (broken power-law continuum and emission line features) are indicated with different colors (see text for details).}
    \label{fig:fit_examples}
\end{figure}

\subsection{AGN spectral fitting} \label{sec:fantasy}

For the modeling of AGN spectra, here we rely on the open-source code \texttt{fantasy} (Fully Automated pythoN Tool for AGN Spectra analYsis)\footnote{\url{https://fantasy-agn.readthedocs.io}}. This is a python-based code for multi-component spectral fitting, optimized for type 1 AGN spectra in the wavelength range 3700-11000 \AA, already successfully used in several studies \citep[][]{ilic20,rakic22,2023A&A...669A.140P}. The AGN spectra are modeled simultaneously with the underlying broken power-law continuum,  the predefined emission line lists, and the Fe II model.
The code is flexible in the selection of different groups of lines, either already predefined lists (e.g., standard narrow lines, Hydrogen and Helium lines, Fe II model, etc), but gives full flexibility to the user to merge predefined line lists or create customized line list. 
Fitting is based on Levenberg–Marquardt algorithm implemented through \texttt{sherpa}\footnote{\url{https://pypi.org/project/sherpa/}} python package \citep[][]{doug_burke_2022_7186379}. 
We describe below the most important features of the \texttt{fantasy} code used in this analysis:
\begin{enumerate}
    \item Several pre-processing steps to prepare the AGN spectra for multi-component fitting are available, such as the Galactic extinction and cosmological redshift correction. Based on either data provided in the header or manually inserted, the spectra are corrected for Galactic extinction using dust map data from \cite{schlegel98}, and for the cosmological redshift. 
\item To estimate and subtract the contribution of the host-galaxy starlight, we have decided for the approach already tested and used for SDSS spectra \citep[][]{2020ApJS..249...17R}, which shows that most AGN spectra can be reconstructed as a linear combination of galaxy and quasar eigenspectra. Using the Principle Component Analysis, \cite{yip04a,yip04b} constructed from 170,000 galaxy SDSS spectra and 16,707 quasar SDSS spectra, a set of galaxy and quasar eigenspectra. \cite{vanden_berk06} showed that {the} majority of AGN spectra can be reconstructed using the linear combination of 10 quasar eigenspectra from \cite{yip04b} and 5 galaxy eigenspectra from \cite{yip04a} as \begin{equation}
F(\lambda)=\sum_0^{10}a_iq_i(\lambda)+\sum_0^5b_ih_i(\lambda)
\end{equation}
where $F(\lambda)$ is observed spectra, $a_i$, and $b_i$ are linear coefficients and $q_i$ and $h_i$ are quasar and galaxy eigenspectra, respectively. Note that prior to the fitting, the observed and eignespectra are binned to the same spectral resolution and wavelength range.
Optionally, \texttt{fantasy} may use all available eigenvectors \citep[10 eignevectors for galaxy (stellar) and 15 for quasar components,][]{yip04a,yip04b}. In this case the code will test for different number of components until reaching {the} best result based on the $\chi^2$ parameter. In both cases, the weighted fit is used in order to avoid accounting for strong emission lines, with an option to mask strong narrow emission lines. By subtracting the reconstructed host galaxy contribution from the observed spectrum, one can obtain the pure AGN spectrum. The proposed technique allows for the recovery of the host galaxy spectrum, which is not resolved otherwise, which enables some studies of host galaxy properties, such as type, luminosity, colors, stellar mass, star-formation rates, etc. \citep[see][]{vanden_berk06}.

    \item One challenging step in bulk fitting of AGN spectra with different spectral features is the identification of emission lines and features present. \texttt{fantasy} approaches this by creating predefined standard lists of AGN emission lines within the specified wavelength range, such as: Hydrogen lines (Balmer and Paschen series), Helium lines (both He I and He II), most present narrow emission lines ([O III], [N II]), other AGN narrow lines (e.g., [S II], [O I]), other AGN broad lines (Ca I, O I), coronal lines (e.g. [Fe X], [Ar V]), Fe II model (described in Section 3.1), etc. In case of very strong Fe II emitters, such as NLSy1 galaxies, sometimes it is necessary to include additional transitions from forbidden Fe II \citep[see e.g.,][]{2004A&A...417..515V}. 
Users can modify available line lists, as well as create completely new ones. {All identified line lists are available (Table \ref{tab:lines} in the Appendix) for completeness}, as we found that not much on this aspect is provided in the literature. 
    \item Many works use a standard approach and fit the optical continuum in AGN with a single power-law with adjustable spectral index \citep[e.g.][]{2020ApJS..249...17R}. However, in order to be able to simultaneously fit the {continuum} and emission line features in a wider range of wavelengths (i.e., to cover both H$\alpha$ and H$\beta$ lines), we have decided to go for more flexibility and use broken power-law \citep[][]{2008MNRAS.383..581D}. \cite{2001AJ....122..549V} detected in a composite SDSS spectrum an abrupt change in the continuum slope redward from H$\beta$ line (see their Figure 5), and discussed that the stellar light from the host galaxies may cause the steepening of the spectral index beyond 5000 \AA, but also pointed that it could be a real change in the quasar continuum, caused by the tail-end of thermal emission from hot dust. \cite{2019ApJS..243...21L} demonstrated that for SDSS type 1 AGN, the broken power-law with a break wavelength of $\sim$5650 \AA, is well suited for {a} simultaneous fit of continuum and emission lines. The break wavelength of 5650 \AA\, is adopted because it can ideally avoid the wavelength regions of the prominent emission lines. Therefore, the \texttt{fantasy} code {uses a} more generalized approach with the broken power-law for representing the AGN continuum, with {the} option to define the break wavelength, depending {on} the wavelength range of interest.
    \item Final step is the spectral model construction and fitting. Basic model for AGN spectra should contain an underlying continuum (generalized to be in {the} form of a broken power-law) and narrow and broad emission lines. Depending on the wavelength range, spectral quality (S/N ratio, spectra resolution), and object type, the fitting model can be customized to contain many and complex emission features. Special feature in \texttt{fantasy}  is a possibility to create a ``fixed model'', which calls all lines from the indicated line list(s) and {sets} them to have the same width and shift. An option to create a ``tide model'' includes all lines in {indicated} lists to have width and shift tided to the reference line (this is typically strong narrow [O III] $\lambda$5007 line). {Creating a} ``feii model'' calls for the Fe II model, in which all iron lines have the same width and shift, and line intensity ratios are calculated as described in Section 3.1. All emission lines are modeled with Gaussian function, defined with shift, width, and intensity.
    \item The uncertainties in the spectra, and consequently in measured spectral quantities were estimated using a Monte Carlo approach \citep[see e.g. description in][]{2020ApJS..249...17R}. We created 50 mock {spectra} for each object in the sample, by adding Gaussian random noise to the original spectrum at each pixel. The same fitting model was applied to all mock spectra as was done for the original one. Spectral quantities of interest (flux, line widths and shifts) were estimated from the original and mock spectra, giving us the distribution of each spectral quantity. For the uncertainty we {then take} the semi-amplitude of the range enclosing the 16th and 84th percentiles of the distribution.
\end{enumerate}

\subsubsection{Case of I Zw 1}

To illustrate the potential of the \texttt{fantasy}  code, the multi-component fitting was done on the I Zw 1 observed spectrum in the $\lambda\lambda$4000-6800 \AA\, wavelength region (Fig. \ref{fig:fitZw}), {which was corrected for Galactic extinction}. We note that the I Zw 1 spectrum is pretty flat in the blue part, indicating that there might be significant intrinsic extinction by the host-galaxy \citep[see][]{2006ApJS..166..470R, 2022ApJS..258...38P}.  We {have} tested the host galaxy subtraction provided in \texttt{fantasy} (see Section 3.2), and seen no noticeable difference in the corrected spectrum in the region of interest (near H$\alpha$ and H$\beta$ linea), and consequently in Fe II fittings, so we continued the analysis on the I Zw 1 spectrum not corrected for the host-galaxy contribution. The model (red line) consists of: broken power-law (black dashed line), narrow lines (green solid line), broad (blue solid line) and intermediate (blue dashed line) components of Balmer lines: H$\alpha$, H$\beta$, H$\delta$, He I lines (yellow line), intermediate components of [O III] (blue dashed line), and Fe II model (dark-red line) and [Fe II] lines (light-red line). All lines within each modelling component are set to have the same width and shifts. We note here that since I Zw 1 and other NLSy1 are known to have strong forbidden ion emission, the modeling also includes these lines, set to have the same width and shift as Fe II model. The list of [Fe II] {lines is compiled from \cite{2004A&A...417..515V} (Table \ref{tab:lines} in the Appendix)}. The model describes remarkably the observed spectrum, as seen through the residual spectrum (bottom panel, Fig. \ref{fig:fitZw}).

\subsubsection{Sample of SDSS AGN spectra}

We prepared the 655 spectra from the SDSS sample ({through procedures listed in} step 1) and subtracted the reconstructed host galaxy contribution (obtained through step 2) from the observed spectrum (see an example in Fig. \ref{fig:host}). On the shown example, the host-galaxy {contributes} on the level of 76\% to the observed continuum flux around 6000 \AA. In {a} few cases when stellar contribution from the host galaxy is estimated to be below zero, it has not been subtracted from the observed spectrum. We then performed simultaneous multi-component spectral fitting with the \texttt{fantasy} code, aiming to measure the fluxes and widths of the pure broad component of Fe II, H$\gamma$, H$\beta$ and H$\alpha$ lines. 

 We fitted the bulk of the spectra in the rest wavelength range $\sim$4000-7000 \AA\ using a single model\footnote{Fittings were run on the SUPERAST computer cluster of {the} Department of Astronomy, University of Belgrade - Faculty of Mathematics \citep[][]{2022PASRB..22..231K}.} consisting {of}: i) broken power-law continuum, to allow for the simultaneous fitting of {the} H$\alpha$ and H$\beta$ wavelength range, which could have {a} different continuum slope \citep[][]{2001AJ....122..549V}; the breaking point was set to be in the range 5350--5650 \AA, which is free from strong emission lines; ii) broad hydrogen (H$\alpha$, H$\beta$, H$\gamma$, H$\delta$) and helium (He I 5877$\mathrm{\mathring{A}}$, He II 4686$\mathrm{\mathring{A}}$) lines; iii) very broad component for strong hydrogen lines (H$\alpha$, H$\beta$, H$\gamma$) lines; iv) standard strong narrow emission lines, all fixed to have the same shifts and widths as [O III] 5007$\mathrm{\mathring{A}}$: [O III] 4363$\mathrm{\mathring{A}}$, [O III] $\lambda\lambda$4959, 5007$\mathrm{\mathring{A}}$, [N II] $\lambda\lambda$6548, 6583$\mathrm{\mathring{A}}$, [S II] $\lambda\lambda$6716, 6731$\mathrm{\mathring{A}}$, [O I] $\lambda\lambda$6300, 6364$\mathrm{\mathring{A}}$; the ratios of [O III] and [N II] doublets were fixed to 3 \citep[][]{2007MNRAS.374.1181D,2022A&A...659A.130K,2023AdSpR..71.1219D};  v) the broad component of the [O III] doublet, which line ratio is also fixed to 3, and have same width and shift \citep[][]{2022A&A...659A.130K}; vi) optical Fe II model, described in Section 3.1, in which all lines have same width and shift. 
The $\chi^2$ was used to test for the goodness of the fitting results, however all results were also visually inspected.

\begin{figure*}[ht!]
    \centering
    \includegraphics[width=0.9\textwidth]{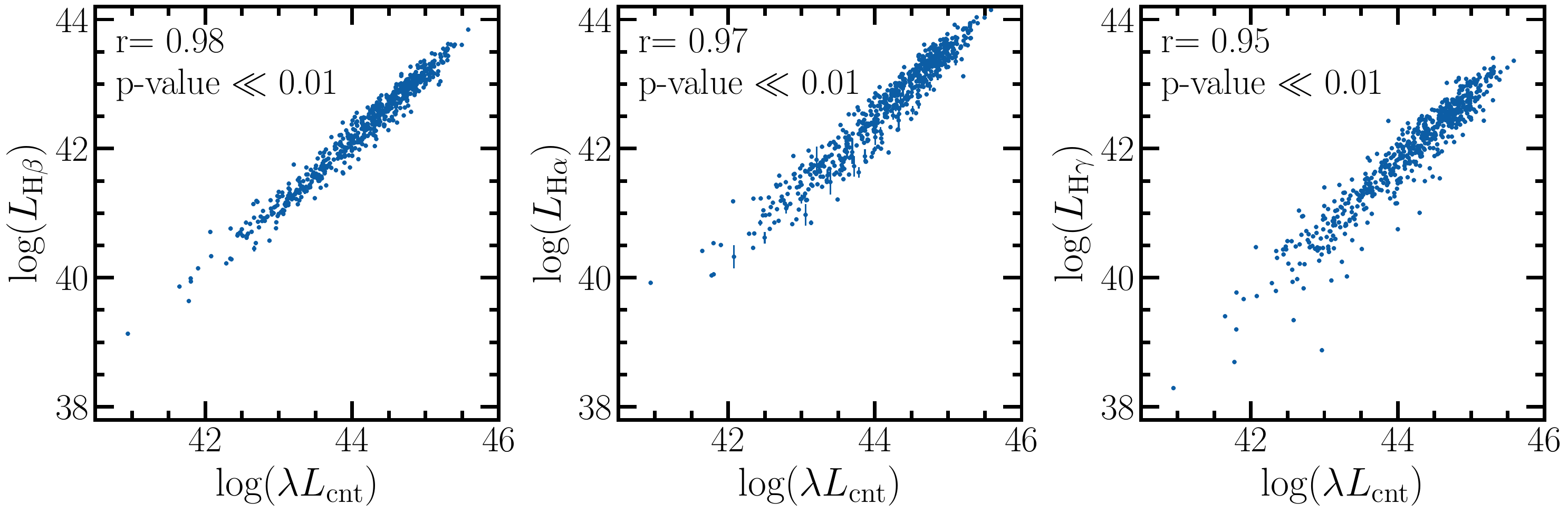}
    \includegraphics[width=0.9\textwidth]{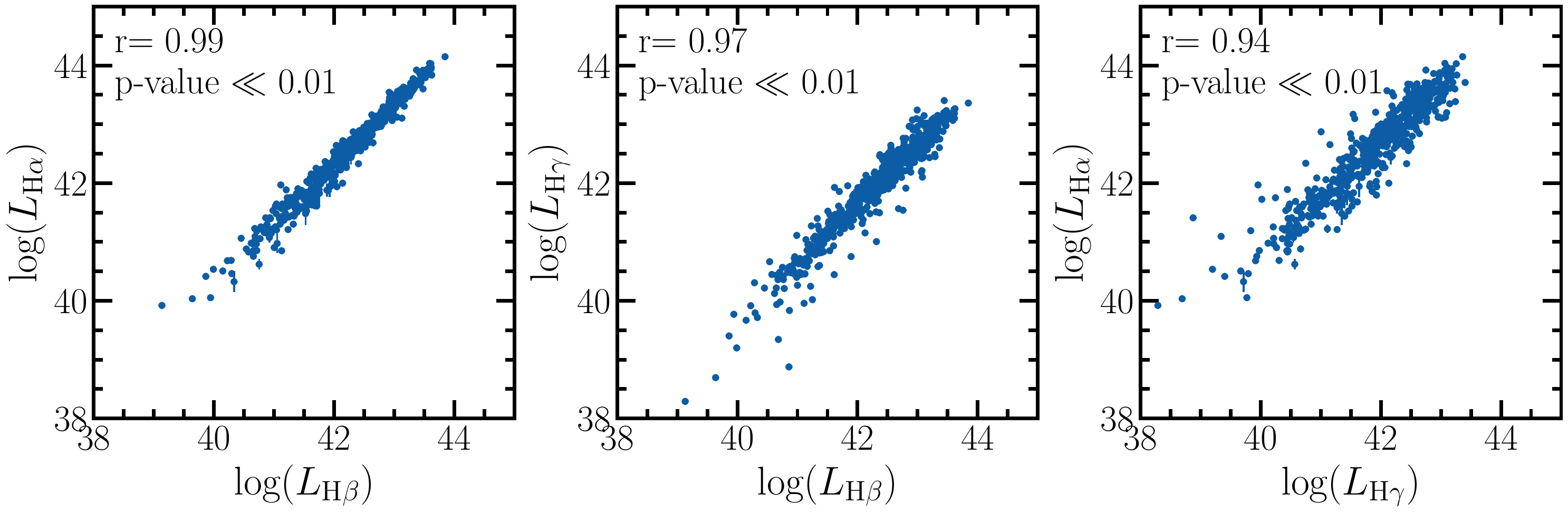}
    \caption{Luminosity (in erg s$^{-1}$) of the broad H$\alpha$ (left), H$\beta$ (middle), and H$\gamma$ (right) line as a function of the continuum luminosity $\lambda L_{\rm cnt}$\AA\, at 5100 \AA\, (upper panels) and for  
    broad H$\alpha$ vs. H$\beta$ (left), H$\alpha$ vs. H$\gamma$ (middle), and H$\beta$ vs. H$\gamma$ (right) for the whole SDSS sample (bottom panles). Pearson correlation coefficient together with corresponding p-value is indicated on each plot. }
    \label{fig:lum}
\end{figure*}

\begin{figure*}[ht!]
    \centering
    \includegraphics[width=0.9\textwidth]{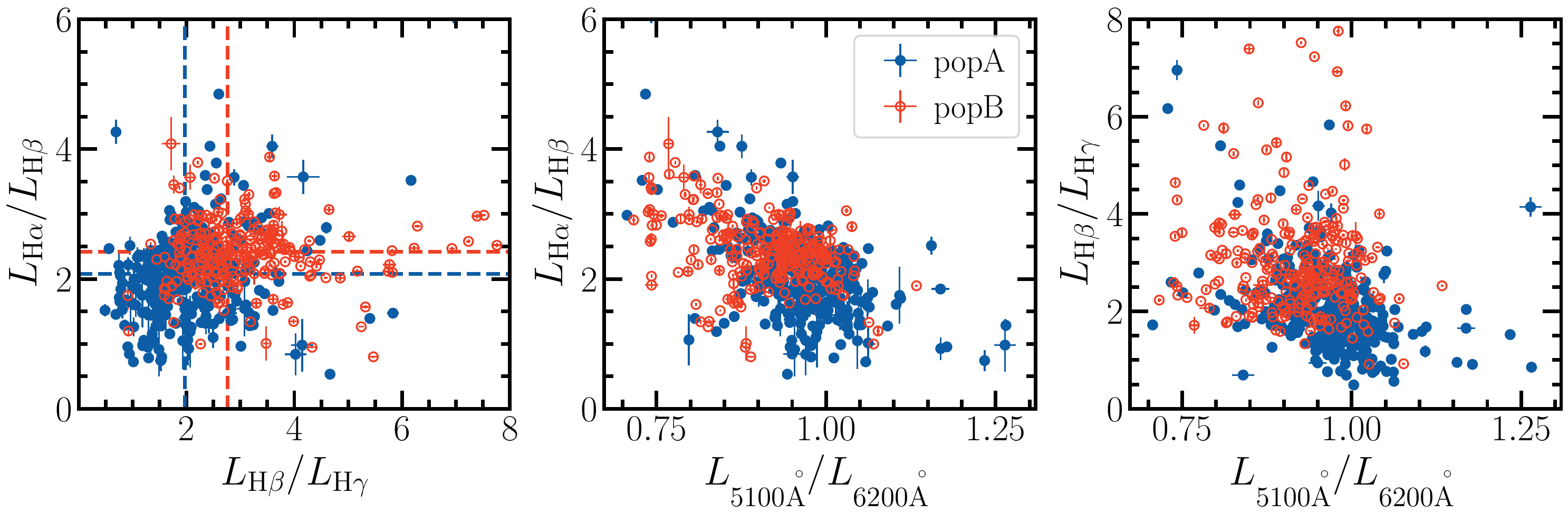}
    \caption{Balmer decrement H$\alpha$/H$\beta$ vs. H$\beta$/H$\gamma$ (left) for pop A (blue) and pop B (red) sub-samples. Middle and right panel show H$\alpha$/H$\beta$ and H$\beta$/H$\gamma$  vs. ratio continuum luminosities at 5100 \AA\, and 6200 \AA. Significant anti-correlation is seen with the continuum ratio: for pop A object r$=-0.51$ ($p_0\ll$0.01) for H$\alpha$/H$\beta$ and r$=-0.40$ ($p_0\ll$0.01) for  H$\beta$/H$\gamma$, and for pop B object  r$\sim-0.39$ ($p_0\ll$0.01) for H$\alpha$/H$\beta$ and r$=-0.25$ ($p_0\ll$0.01) for  H$\beta$/H$\gamma$.}
    \label{fig:decrement}
\end{figure*}

\begin{figure*}[ht!]
    \centering
    \includegraphics[width=0.6\textwidth]{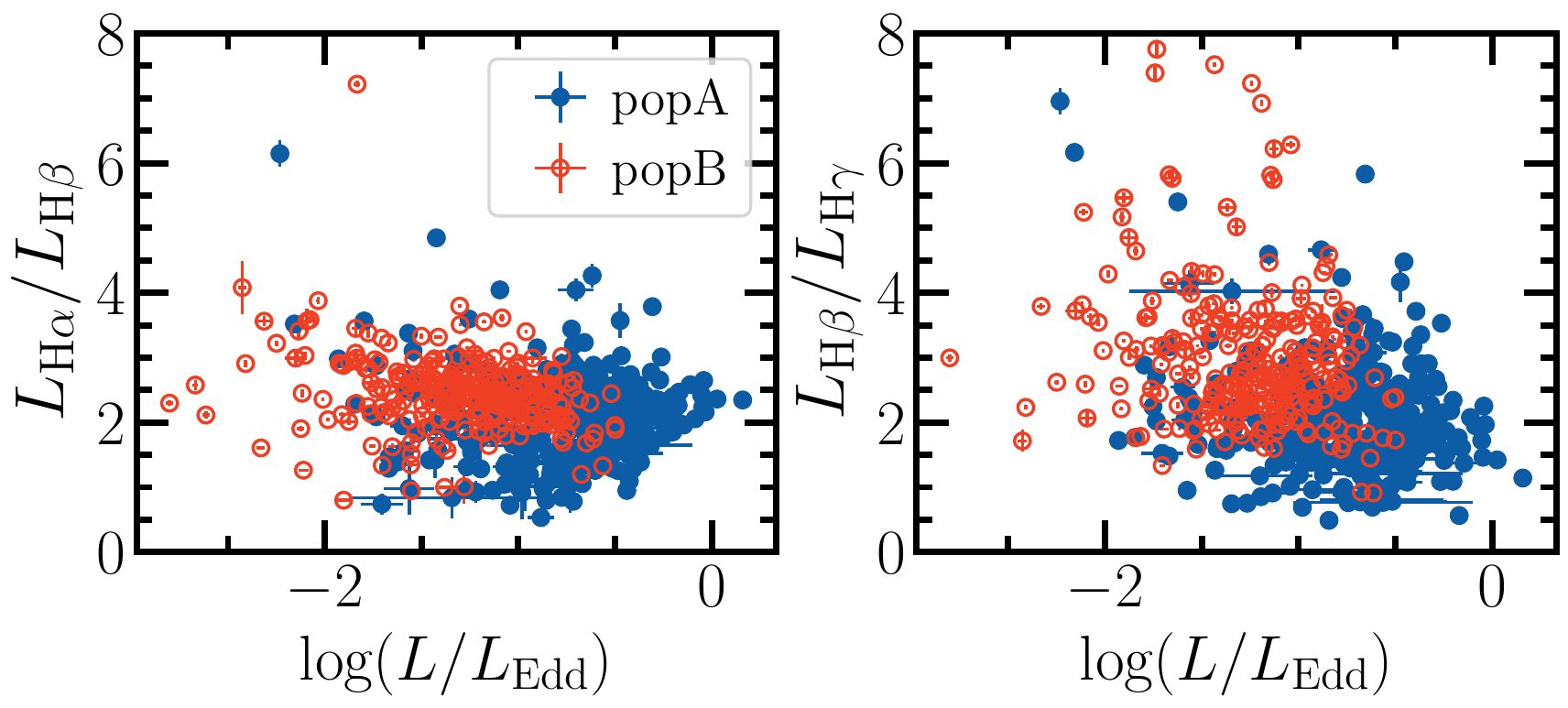}
    \caption{Balmer decrement H$\alpha$/H$\beta$ (left) and H$\beta$/H$\gamma$ (right) for pop A (blue) and pop B (red) sub-samples vs. Eddington ratio ($L/L_{\rm Edd}$). Pearson correlation coefficients point to no correlations: for pop A object r$=0.05$ ($p_0=0.3$) for H$\alpha$/H$\beta$ and r$=-0.22$ ($p_0<$0.01)} for  H$\beta$/H$\gamma$, and for pop B object r$\sim-0.24$ ($p_0<$0.01) for H$\alpha$/H$\beta$ and r$\sim-0.16$ ($p_0=$0.01) for  H$\beta$/H$\gamma$. 
    \label{fig:decrement_edd}
\end{figure*}

\section{Results and discussion} \label{subsec:results}

The sample of 655 SDSS spectra were fitted with the single spectral model defined in the previous section ({a} few examples are given in Fig. \ref{fig:fit_examples}). After visual inspection, 34 objects ($\sim$5\%) {were} excluded from further analysis, mostly due to lower S/N or very broad or double-peaked profiles, which could not be addressed with the single model described here. We have noticed that in {a} few cases, the underlying stellar continuum was well reproduced (based on the galaxy absorption features), and subtracted, however the pure AGN spectra {were} still showing {a} slight increase towards larger wavelength (as illustrated in the bottom panel of Figure \ref{fig:fit_examples}). This continuum reddening could be interpreted either as a result of poor removal of the host-galaxy stellar continuum using the method of spectral decomposition with galaxy templates, or possibly we see traces of {the} contribution coming from a tail-end of thermal emission from the hot dust present within the AGN.

{After decomposing the spectra into individual components, we} measured the following spectral parameters from the best-fit model: broad H$\alpha$, H$\beta$, and H$\delta$ fluxes, continuum luminosities ($\lambda$L$_\lambda$) at 5100\AA\, and 6200 \AA\, (median integral in 5090--5110 \AA \, and 6190--6210 \AA\, respectively, from the reconstructed pure AGN spectra, Figure \ref{fig:host}), and Fe II in three different windows: Fe II blue (4340--4680 \AA), Fe II green (5100--5600 \AA), and Fe II red (6100--6650 \AA). Fluxes were measured from the modeled broad line profiles and continuum, and then converted to luminosities based on the luminosity distance calculated from the cosmological redshift and adopted cosmological parameters (see Section 1).  The Fe II blue band  was used to get the $R_{\rm Fe II}$ parameter. The FWHM of the lines is also measured, and in case of broad Hydrogen lines, which were fitted with two Gaussians, the width of the total broad line was calculated. From the measured continuum luminosity at 5100 \AA\, and FWHM of H$\beta$ line, we get the $M_{\rm BH}$ through standard single-epoch method for SMBH mass estimates \citep[see e.g.,][]{pop20,DallaBonta20}. 
Once we have $M_{\rm BH}$, the Eddington luminosity is simply 
$L_\mathrm{Edd} = 1.26\times 10^{38}(M_\mathrm{BH}/M\odot)$ erg s$^{-1}$. For the bolometric luminosity  $L_{\rm bol} = k_{\rm bol} \lambda L_\lambda$ we used the mean quasar bolometric correction $k_{\rm bol} \approx 10$ \citep[e.g.,][]{2006ApJS..166..470R} and the continuum luminosity at 5100 \AA. {This then gives} the Eddington ratio $L_{\rm Edd}/L_{\rm bol}$. The uncertainties of measured quantities are calculated as described in Section 3.2 (step 6), and then further propagated for derived quantities like luminosity or Eddington ratio. Some measured quantities, especially of strong broad emission lines (e.g. fluxes of H$\alpha$ and H$\beta$ lines or Fe II lines in the sample of xA objects) have low uncertainties, which is a result of spectra being selected to have higher S/N$ >$35. For cases when we could not estimate the uncertainties, we used the mean value [in \%] to get the uncertainty of the measured quantity.

Table \ref{tab:results} lists the measured spectral parameters with uncertainties, that is: the SDSS object ID, redshift, broken power-law indices of the fitted underlying AGN continuum ($\alpha_1$,$\alpha_2$ where $\alpha_1$ describes the part of spectra with wavelengths larger than the break wavelength, and $\alpha_{1}+\alpha_{2}$ the lower ones), continuum luminosity L$_{5100}$ and L$_{6200}$,  the H$\alpha$, H$\beta$ broad line luminosities, as well as the luminosities of Fe II blue, green, red, the full width half {a} maximum of the broad H$\alpha$, broad H$\beta$, and Fe II lines, and Eddington ratio $L/L_{\rm Edd}$. The last row gives the mean average quantities for the total SDSS sample. 
{ Table \ref{tab:results} is available in its entirety in the machine-readable format in the online Journal.}

\subsection{Hydrogen Balmer lines}

First we present our results for the hydrogen Balmer lines in the total sample. Figure \ref{fig:lum} (upper panels) shows the strong correlation of the luminosity of broad Balmer lines  (H$\alpha$, H$\beta$,  H$\gamma$) as a function of the continuum luminosity $L_{5100}$, for the whole SDSS sample. High Pearson correlation coefficients close to unity support this. Strong correlations between line and continuum luminosities have been observed before by many studies in both single object \citep[e.g.,][]{2017FrASS...4...12I,DallaBonta20} and larger samples \citep[e.g.,][]{2019ApJS..243...21L,2020ApJS..249...17R}, supporting that the results obtained through the \texttt{fantasy} fittings are in agreement with previous findings. The observed line-continuum correlations are expected if photoionization by the central continuum emission is the main heating source of the BLR and thus responsible for the broad line emission \citep[][]{2006agna.book.....O,2013peag.book.....N}. In case of H$\gamma$ line (Fig. \ref{fig:lum}, right panels), there is some scatter for lower line luminosities, probably due to difficulties to identify and subtract satellite lines such as [O III] $\lambda$4363. 

It is important to know that the intrinsic extinction due to {the}  presence of dust within the AGN makes it unclear what fraction of luminosity is actually being measured \citep{kaspi2000}. This AGN intrinsic reddening is still one of the critical {points} in the studies of the AGN phenomenon \citep{2017MNRAS.467..226G}. Dust is present not just in the host galaxy, but is also associated with the central regions of the AGN, {and} not just in the equatorial part typically {explained} by the unified model of AGN, but also in parsec-scale polar areas \citep{2012ApJ...755..149H,2017MNRAS.472.3854S}. In this analysis we have preformed only the correction for the host galaxy contribution, with which we assumed that the extinction within the galaxy (if present) has been removed. However, the effects of extinction within the AGN itself have not been assessed \citep[as not in other studies, e.g.][]{2017MNRAS.472.4051C,2019ApJS..243...21L,2020ApJS..249...17R}. Therefore, some contamination to the continuum luminosity may still exist due to intrinsic AGN extinction. The obtained strong luminosity correlations may point to the internal extinction being not strong in most type 1 AGN studied here, as pointed before by some authors \citep[e.g.,][]{2017MNRAS.472.4051C}.

Strong correlations between different broad Balmer line luminosities (H$\alpha$ vs. H$\beta$, H$\alpha$ vs. H$\gamma$, and H$\beta$ vs. H$\gamma$) for the whole SDSS sample are also present (Figure \ref{fig:lum}, bottom panels). Especially in case of H$\alpha$ vs. H$\beta$ (r=0.99 for the total SDSS sample), implying that the emission lines have the same physical origin. This is supported with the same kinematics of these lines, as they have {the} same FWHM, as also shown in \cite{rakic22}. No difference is seen in above correlations when considering different populations, i.e. for pop A, pop B and xA sub-samples.

\movetabledown=2.5in
\begin{rotatetable}
\begin{deluxetable}{lccccccccccccc}
\tabletypesize{\tiny}
\tablewidth{0pt} 
\tablecaption{Measured spectral quantities for the first 10 objects
    of the total SDSS sample. The last row gives the mean average
    quantities for the total SDSS sample.\label{tab:results}}
\tablehead{\colhead{SDSS Object ID} &
\colhead{redshift} &
\colhead{$\alpha_{1}$,$\alpha_{2}$} &
\colhead{log $\lambda$L$_{5100}$} &
\colhead{log $\lambda$L$_{6200}$} &
\colhead{log L(H$\beta$)} &
\colhead{log L(H$\alpha$)} &
\colhead{log L(Fe II$_{\text{blue}}$)} &
\colhead{log L(Fe II$_{\text{green}}$)} &
\colhead{log L(Fe II$_{\text{red}}$)} &
\colhead{w(H$\beta$)} &
\colhead{w(H$\alpha$)}  &
\colhead{w(Fe II)} &
\colhead{$L/L_{\text{Edd}}$} \\
\colhead{} &
\colhead{} &
\colhead{} &
\colhead{erg/s} &
\colhead{erg/s} &
\colhead{erg/s} &
\colhead{erg/s} &
\colhead{erg/s} &
\colhead{erg/s} &
\colhead{erg/s} &
\colhead{km/s} &
\colhead{km/s} &
\colhead{km/s} &
\colhead{}} 
\startdata 
SDSS J145824.46	&	0.246	&	-2.31,		&	44.580$\pm$0.001	&	44.571$\pm$0.001	&	42.558$\pm$0.006	&	42.717$\pm$0.004	&	42.607$\pm$0.005	&	42.428$\pm$0.009	&	41.508$\pm$0.038	&	2829$\pm$35	&	4556$\pm$69	&	2378$\pm$25	&	0.283$\pm$0.008	\\
+363119.5&&-0.22&&&&&&&&&&&\\
SDSS J095302.64	&	0.273	&	-1.32,		&	44.609$\pm$0.001	&	44.637$\pm$0.001	&	42.888$\pm$0.003	&	43.289$\pm$0.001	&	42.407$\pm$0.020	&	42.253$\pm$0.012	&	42.059$\pm$0.018	&	3177$\pm$35	&	3037$\pm$23	&	3367$\pm$48	&	0.232$\pm$0.005	\\
+380145.2&&-1.04&&&&&&&&&&&\\
SDSS J225603.37	&	0.363	&	-2.11,		&	45.273$\pm$0.001	&	45.266$\pm$0.001	&	43.443$\pm$0.005	&	43.808$\pm$0.001	&	43.172$\pm$0.008	&	42.988$\pm$0.006	&	42.578$\pm$0.018	&	2900$\pm$69	&	2693$\pm$35	&	3451$\pm$50	&	0.580$\pm$0.030	\\
+273209.5&&-1.43&&&&&&&&&&&\\
SDSS J224113.54	&	0.058	&	-1.19,		&	43.135$\pm$0.002	&	43.196$\pm$0.001	&	41.384$\pm$0.003	&	41.756$\pm$0.002	&	40.650$\pm$0.046	&	40.007$\pm$0.104	&	40.256$\pm$0.045	&	6835$\pm$68	&	5729$\pm$1	&	2405$\pm$118	&	0.010$\pm$0.001	\\
-012108.8&&0.94&&&&&&&&&&&\\
SDSS J105007.75	&	0.134	&	-2.03,		&	44.571$\pm$0.001	&	44.566$\pm$0.001	&	42.715$\pm$0.002	&	42.971$\pm$0.003	&	42.332$\pm$0.066	&	42.320$\pm$0.008	&	41.712$\pm$0.019	&	2209$\pm$35	&	2278$\pm$0	&	1963$\pm$32	&	0.459$\pm$0.015	\\
+113228.6&&-1.63&&&&&&&&&&&\\
SDSS J125851.45	&	0.075	&	0.56,		&	43.094$\pm$0.001	&	43.187$\pm$0.001	&	41.163$\pm$0.006	&	41.369$\pm$0.004	&	40.749$\pm$0.088	&	40.305$\pm$0.041	&	37.785$\pm$1.350	&	9661$\pm$242	&	6007$\pm$40	&	1945$\pm$129	&	0.005$\pm$0.000	\\
+235526.6&&-0.15&&&&&&&&&&&\\
SDSS J080131.96	&	0.157	&	-1.83,		&	44.691$\pm$0.001	&	44.695$\pm$0.001	&	42.886$\pm$0.011	&	43.310$\pm$0.002	&	42.514$\pm$0.016	&	42.246$\pm$0.007	&	42.046$\pm$0.013	&	7316$\pm$203	&	6494$\pm$69	&	6000$\pm$5	&	0.048$\pm$0.003	\\
+473616.0&&-0.84&&&&&&&&&&&\\
SDSS J125741.05	&	0.081	&	-1.38,		&	43.150$\pm$0.002	&	43.165$\pm$0.001	&	41.366$\pm$0.008	&	41.534$\pm$0.005	&	41.409$\pm$0.012	&	41.229$\pm$0.011	&	40.401$\pm$0.061	&	1450$\pm$35	&	1105$\pm$34	&	3575$\pm$251	&	0.221$\pm$0.010	\\
+202347.8&&-1.62&&&&&&&&&&&\\
SDSS J004719.39	&	0.039	&	-0.50,		&	43.178$\pm$0.001	&	43.235$\pm$0.001	&	41.215$\pm$0.004	&	41.634$\pm$0.002	&	40.895$\pm$0.010	&	40.253$\pm$0.035	&	40.341$\pm$0.038	&	2623$\pm$34	&	2485$\pm$14	&	2630$\pm$31	&	0.070$\pm$0.002	\\
+144212.6&&-1.19&&&&&&&&&&&\\
SDSS J154348.62	&	0.318	&	-1.55,		&	44.830$\pm$0.002	&	44.855$\pm$0.001	&	43.123$\pm$0.004	&	43.534$\pm$0.003	&	42.568$\pm$0.018	&	42.379$\pm$0.036	&	42.428$\pm$0.019	&	4831$\pm$69	&	4627$\pm$40	&	5419$\pm$158	&	0.128$\pm$0.004	\\
+401324.8&&-0.44&&&&&&&&&&&\\
\hline
Averaged	                &		    &	-1.42,	   &	44.160$\pm$0.001	&	44.187$\pm$0.001	&	42.312$\pm$0.006	&	42.644$\pm$0.010	&	41.944$\pm$0.037	&	41.832$\pm$0.023	&	41.400$\pm$0.124	&	4132$\pm$78	&	3765$\pm$71	&	3218$\pm$132	&	0.164$\pm$0.011	\\
 &&-0.72&&&&&&&&&&&\\
\enddata
\tablecomments{Columns give the SDSS object ID, redshift, indices
    $\alpha_{1}$, $\alpha_{2}$ of the broken power-law fitting the
    underlying AGN continuum, logarithm of the continuum luminosities
    at 5100 \AA\, and 6200 \AA\, ($\lambda$L$_{\lambda}$), logarithm of
    the H$\beta$ and H$\alpha$ broad-line luminosities (L(H$\beta$), 
    L(H$\alpha$)), and luminosities of Fe II blue, green, red (L(Fe
    II$_{\text{blue}}$), L(Fe II$_{\text{green}}$), L(Fe II$_{\text{red}}$),
    the full width half maximum of the broad H$\beta$, broad
    H$\alpha$, and Fe II lines in km/s (w(H$\beta$), w(H$\alpha$),
    w(Fe II)), and the Eddington ratio $L/L_{\text{Edd}}$.
    Table \ref{tab:results} is published in its
    entirety in the machine-readable format. A portion is shown here
    for guidance regarding its form and content.}
\end{deluxetable}
\end{rotatetable}

\begin{figure}
    \includegraphics[width=0.52\columnwidth]{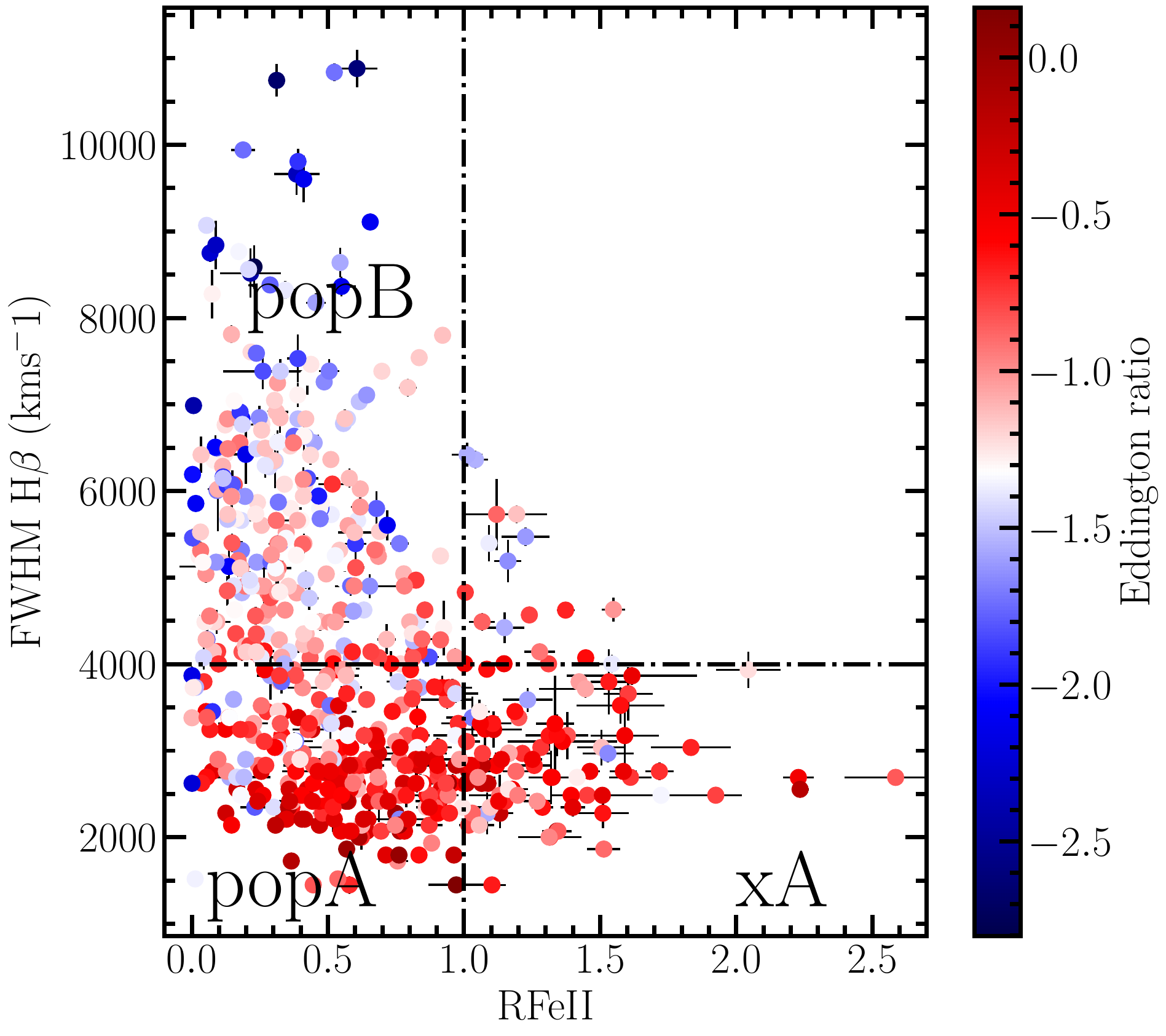}
    \includegraphics[width=0.45\columnwidth]{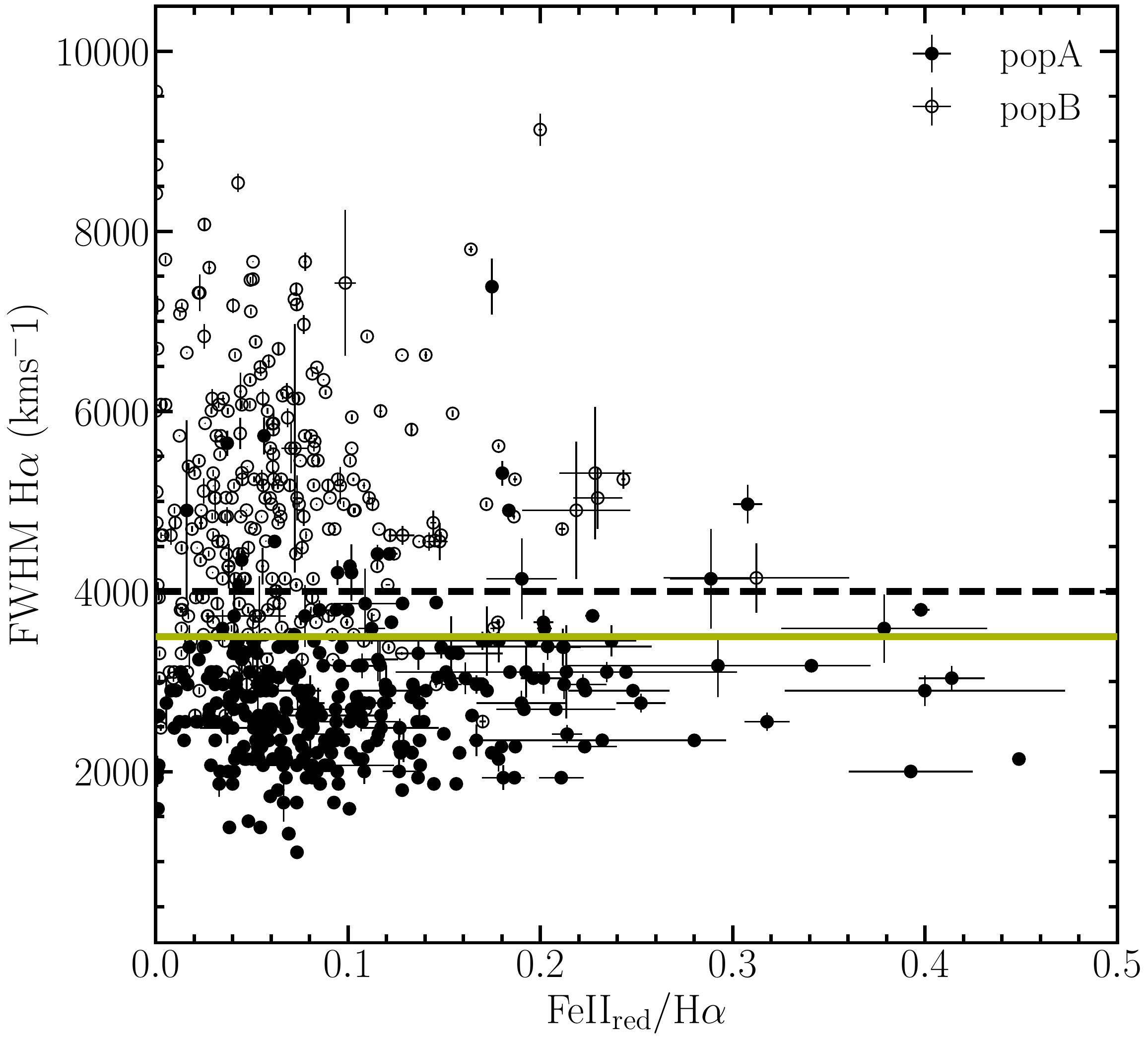}
    \caption{{\it Upper}: SDSS sample plotted on the main sequence diagram (FWHM(H$\beta$) vs. R$_{\rm Fe II}$). Horizontal (FWHM(H$\beta$)=4000 km s$^{-1}$) and vertical (R$_{\rm Fe II}$=1) line divide the population B (pop B), population A (pop A), and extreme population A (xA). Eddington ratio ($L/L_{\rm Edd}$) is shown on the colorbar. {\it Bottom}: Same diagram produced using the H$\alpha$ FWHM and Fe II red (6100--6650 \AA)  emission. Pop A and B identified through the upper main sequence are marked with different symbols. Dashed and solid lines mark the position of FWHM(H$\alpha$)=4000 km s$^{-1}$ and  FWHM(H$\alpha$)=3500 km s$^{-1}$, respectively.}
\label{fig:ms}
\end{figure}

\begin{figure*}[ht!]
    \centering
    \includegraphics[width=0.9\textwidth]{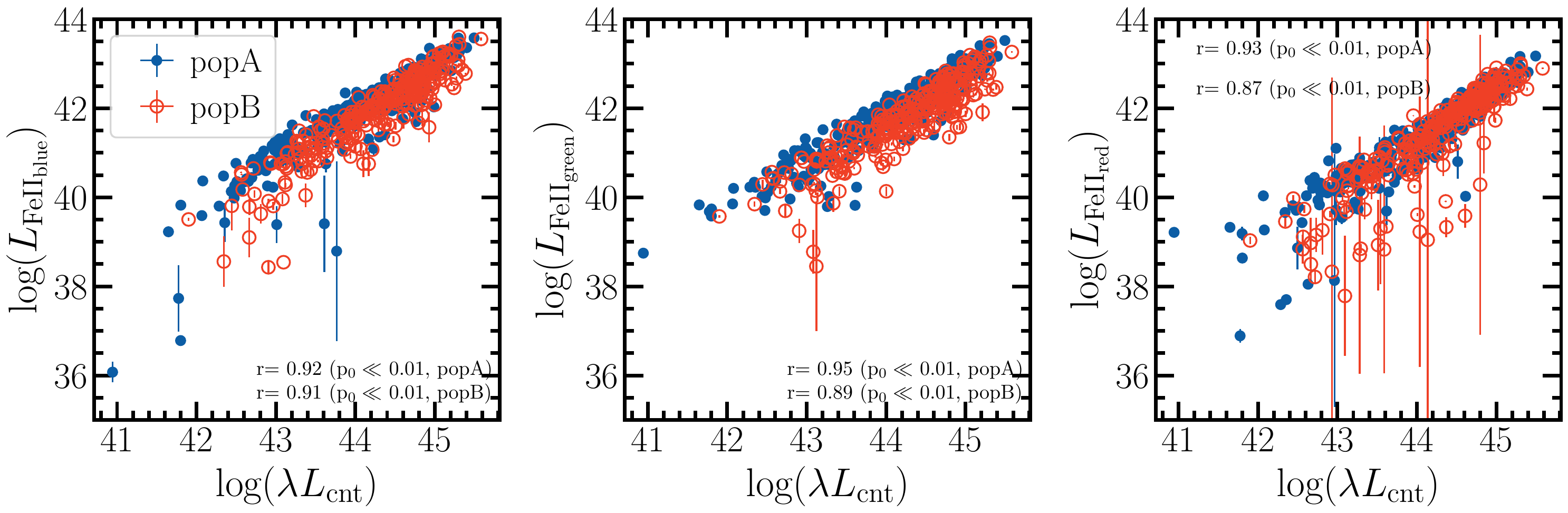}
     \includegraphics[width=0.9\textwidth]{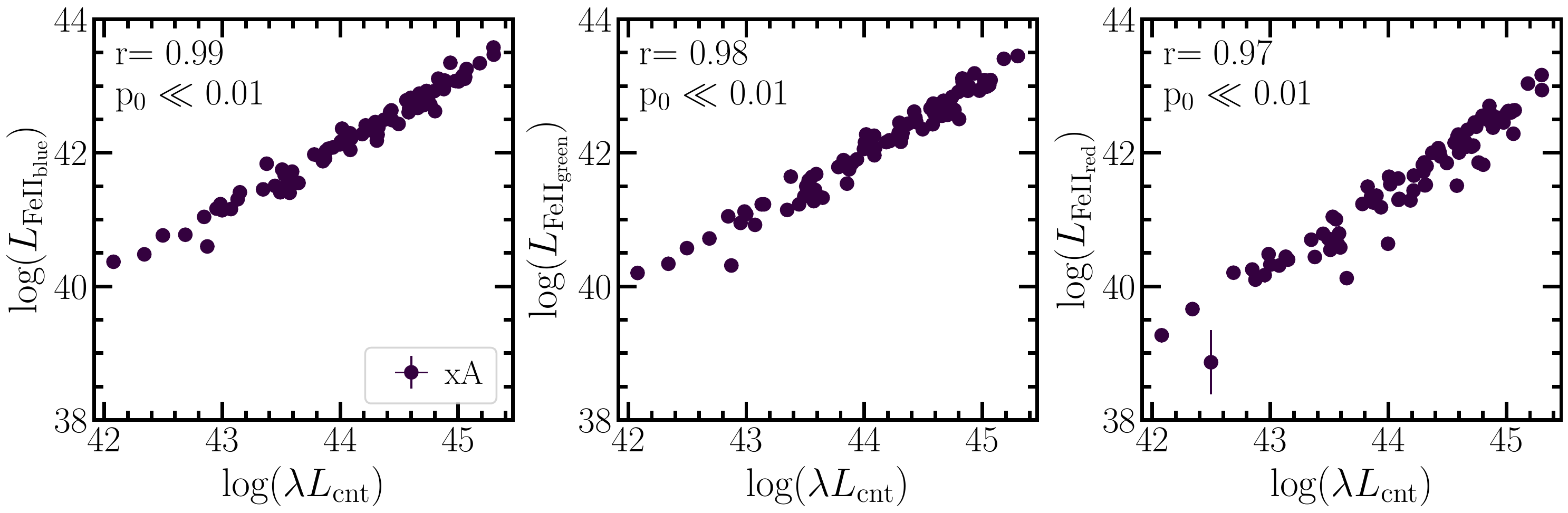}
    \caption{Upper panels show luminosity of Fe II in three bands (Fe II blue (left), Fe II green (middle), and Fe II red (right)) as a function of the continuum luminosity $L_{5100}$\AA\, (given in erg s$^{-1}$) for the pop A (full circles) and pop B (open circles) samples. Bottom panels shows only a sub-sample of xA objects. Pearson correlation coefficient together with corresponding p-value is indicated on each plot.}
    \label{fig:lumfe}
\end{figure*}

The ratio of Balmer lines can tell us about the physical processes of the region where they originate from \citep[e.g.,][]{2007ApJ...671..104L,2012A&A...543A.142I}. 
Balmer decrement H$\alpha$/H$\beta$ vs. H$\beta$/H$\gamma$ is given in Figure \ref{fig:decrement}, in which Pop A and Pop B objects occupy slightly different {areas}. Pop B show higher Balmer decrement (average H$\alpha$/H$\beta$=2.42, H$\beta$/H$\gamma$=2.76) than pop A objects (average H$\alpha$/H$\beta$=2.07, H$\beta$/H$\gamma$=1.97). These values are below those (H$\alpha$/H$\beta\approx 3$) suggested by pure recombination theory. One possibility could be that collisional deexcitation is decreasing the H$\alpha$ line, and thus giving a lower H$\alpha$/H$\beta$ ratio. {It is noteworthy that in our results} H$\alpha$ flux may be systematically lower, since {a} part of the emission is allocated to Fe II lines.

The relationship between the Balmer decrement and the ratio of continuum measured at 5100 \AA\ and 6200 \AA\ can be seen in Fig. \ref{fig:decrement}.  {A} significant low-level anti-correlation {is somewhat} stronger for pop A object (r$=-0.51$ ($p_0\ll$0.01) for H$\alpha$/H$\beta$, r$=-0.40$ ($p_0\ll$0.01) for  H$\beta$/H$\gamma$) than in pop B object (r$\sim-0.39$ ($p_0\ll$0.01) for H$\alpha$/H$\beta$, r$=-0.25$ ($p_0\ll$0.01) for  H$\beta$/H$\gamma$). As the continuum 5100/6200 ratio decreases, meaning the object becomes redder, the Balmer decrement increases. This suggests that the increase of the broad H$\alpha$/H$\beta$ ratio may be due to {the} low-level of reddening, at least in some fraction of object. On the other hand, the values of the Balmer decrement of broad lines H$\alpha$/H$\beta$ measured here are below the theoretical predictions in most objects \citep[][]{2017MNRAS.472.4051C,2019MNRAS.483.1722L}, making it hardly possible to assess extinction using H$\alpha$/H$\beta$ ratio. The possibility that this may however point to some dust presence is supported by previous studies showing that objects classified as ``pop B'' are more affected by dust due to their larger inclination angles, while ``pop A'' objects are seen more face-on and have lower inclination angles \citep{marziani22}, making them less affected by dust. We also {demonstrate} that Balmer decrement {is not dependent} on the Eddington ratio ($L/L_{\rm Edd}$), as shown in Figure \ref{fig:decrement_edd}, supported with no correlation present (correlation coefficients $r\sim 0$). This is in agreement with the previous findings of \cite{2019MNRAS.483.1722L}.

When plotted on the main sequence diagram (FWHM(H$\beta$) vs. R$_{\rm Fe II}$, Fig. \ref{fig:ms}, upper panel), the SDSS sample occupies {the} expected parameter space \citep[][]{2018FrASS...5....6M}. Moreover, the Eddington ratio ($L/L_{\rm Edd}$) gradually {rises} from pop B to pop A sources (colorbar in Fig. \ref{fig:ms}), as expected \citep{2016ApJ...818L..14D}. The average values of $L/L_{\rm Edd}$ are 0.07, 0.24 and 0.23 for pop B, pop A, and xA samples, respectively.  The dependence of the R$_{\rm Fe II}$ sequence on {the} Eddington ratio has been also described through modeling with photoionization codes  \citep{2019ApJ...882...79P}. Measuring the Fe II emission near the H$\alpha$ line, allowed us to construct for {the} first time a main sequence diagram using H$\alpha$ line and fluxes Fe II red (6100--6650 \AA), presented in Fig. \ref{fig:ms}, bottom panel). The same trend with the Eddington ratio increasing from pop B toward pop A objects is also detected. Different populations identified through the standard main sequence diagram (Fig. \ref{fig:ms}, upper panel) occupy similar areas of H$\alpha$ widths and iron strength, only that Fe II (red) emission is much weaker than H$\alpha$ line. We note that the main sequence diagram for H$\alpha$ line points that the division line between pop A and pop B objects may be lower (FWHM (H$\alpha$) $\sim$ 3500 km s$^{-1}$), as indicated with {a} solid horizontal line in Fig \ref{fig:ms}, bottom panel.

\begin{figure*}
    \centering
    \includegraphics[width=0.9\textwidth]{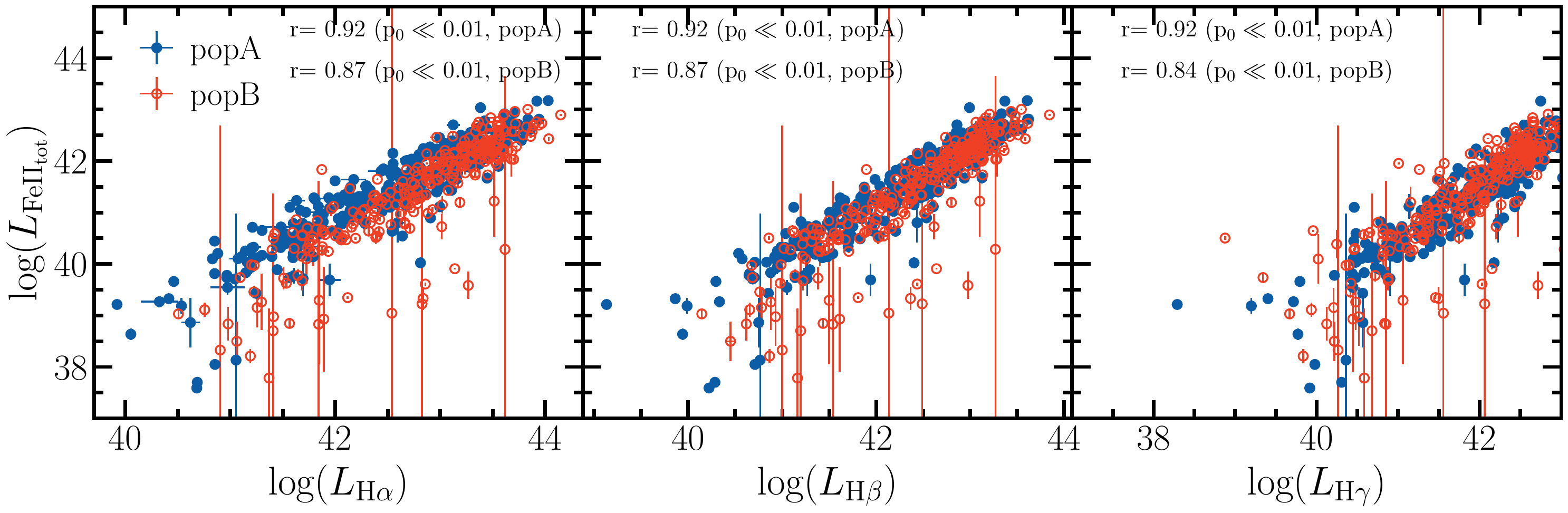}
    \caption{Total Fe II emission in pop A (full blue circles) and pop B (open red circles) samples with respect to the H$\alpha$ (left) and H$\beta$ (right) {luminosities}, in units erg s$^{-1}$. Pearson correlation coefficient together with corresponding p-value is  indicated on each plot. }
    \label{fig:fe_lines}
\end{figure*}

\begin{figure*}
    \centering
    \includegraphics[width=0.7\textwidth]{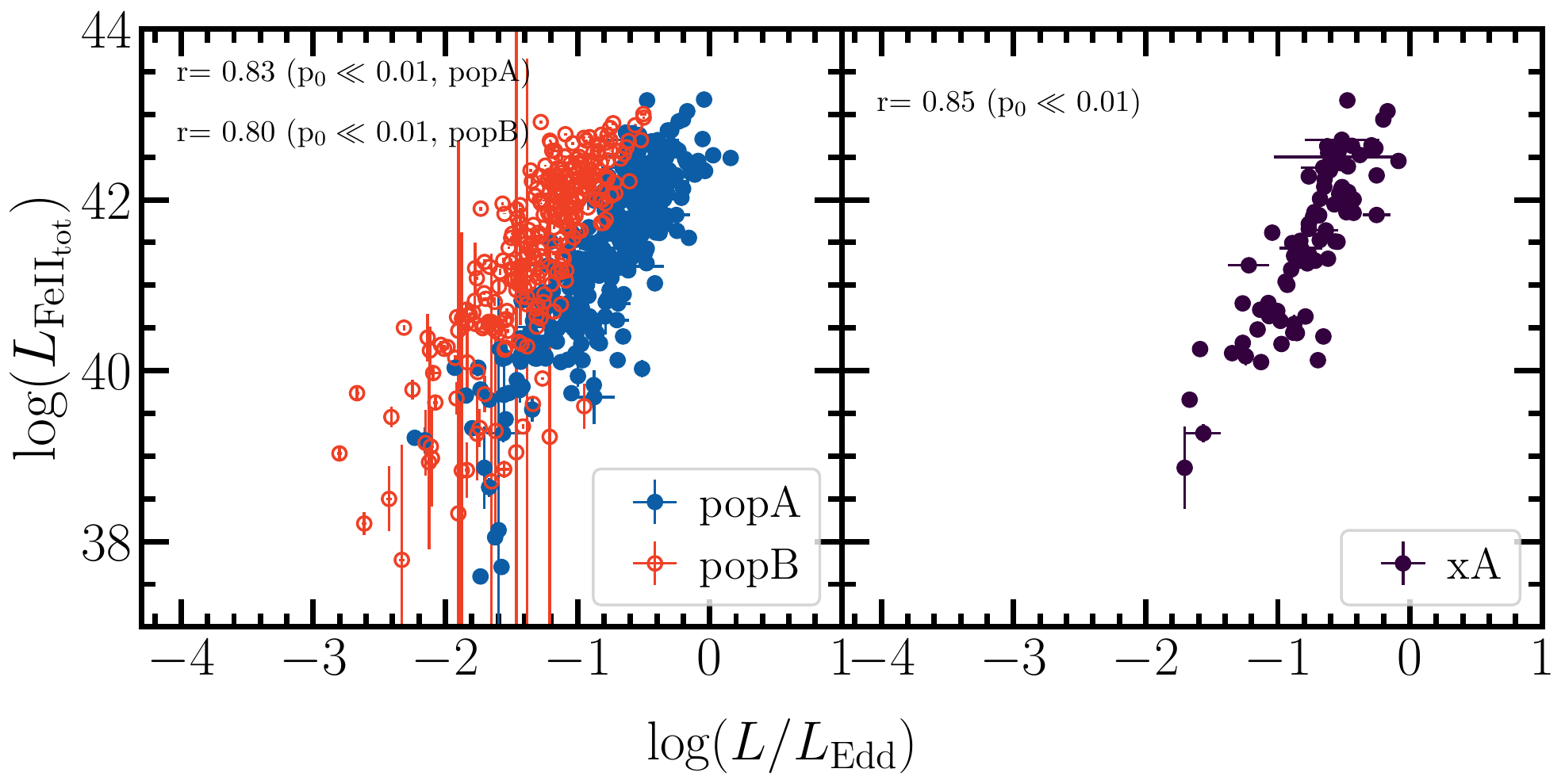}
    \caption{Total Fe II emission in pop A and pop B sample (left), and for xA sample (right) with respect to the Eddington ratio $L_{\rm Edd}/L_{\rm bol}$.}
    \label{fig:fe_edd}
\end{figure*}

\subsection{Fe II emission near H$\alpha$ and H$\beta$ wavelength bands}

More compelling is to explore the behaviour of Fe II emission with respect to other spectral features. Here we study total Fe II emission and in three selected wavelength bands (Fe II blue 4340--4680 \AA, Fe II green 5100--5600 \AA, and Fe II red 6100--6650 \AA) in three populations of type 1 AGN, i.e. in pop A, pop B and xA sub-samples.

In Figure \ref{fig:lumfe} (upper panels), we plot the luminosity of Fe II in three bands: Fe II blue (left), Fe II green (middle), and Fe II red (right) as a function of the continuum luminosity $L_{5100}$\AA\, for the pop A (full circles) and pop B (open circles) samples. Significant correlation is seen for all three studied iron bands, pointing to the importance of the central continuum emission for the Fe II line production. It seems the scatter is slightly larger for pop B objects, however significant {correlation} is present for both sub-samples (correlation coefficient indicated in Figure \ref{fig:lumfe}, upper panels). Similar trends have been observed before by \cite{2011ApJ...736...86D}, but they report {a} lower level of correlations and larger scatter probably due to much lower quality of {the} studied sample.
We note that in case of objects with weaker Fe II emission (or spectra of low S/N), it is more difficult to detect Fe II lines as they blend with the continuum.
This is probably the reason behind the scatter seen in Fe II (red) emission in pop B objects (see right panels in Fig. \ref{fig:lumfe}). In case of xA sample, the correlation with the continuum luminosity {is} even stronger for all three iron bands (Fig. \ref{fig:lumfe}, bottom panels).

\begin{figure*}
    \centering
    \includegraphics[width=0.9\textwidth]{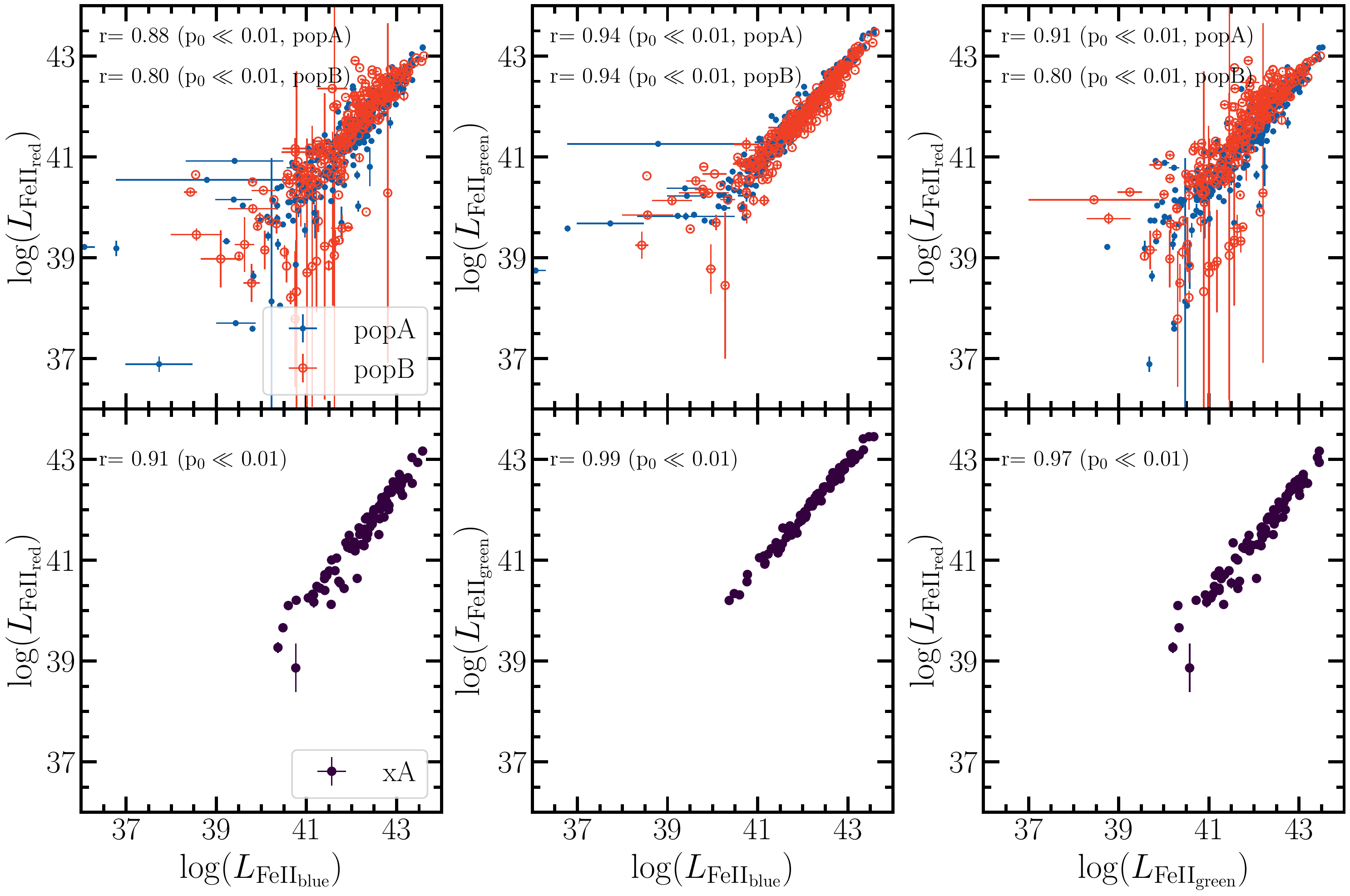}
    \caption{Correlations between luminosity of Fe II in three bands: Fe II blue (4340--4680 \AA), Fe II green (5100--5600 \AA), and Fe II red (6100--6650 \AA) for the pop A (blue circles) and pop B (red circles) samples (upper panels), and only for the population of extreme A objects (bottom panels).  Pearson correlation coefficient together with corresponding p-value is indicated on each plot.}
    \label{fig:fe}
\end{figure*}

Detailed studies of the Fe II emission, including modeling and observations point to complex excitation mechanisms involved in its production. {In their theoretical calculations of Fe II emission, \cite{2003ApJS..145...15S} used for} the excitation mechanisms: continuum and line (Ly$\alpha$, Ly$\beta$) fluorescence, collisional excitation, and self-fluorescence among the Fe II transitions. Some other studies show that collisional excitation is {an} important driver of the Fe II optical emission, with Ly$\alpha$ fluorescence contributing on the level of $\sim$20\% \citep{2004ApJ...615..610B,2012ApJ...751....7G,2016ApJ...820..116M}. Some evidence for photoionization by the central source as responsible for the Fe II emission comes from variability studies. For example, \cite{2013ApJ...769..128B} showed that Fe II emission in two Seyfert 1 galaxies, NGC 4593 and Mrk 1511, does reverberate on short timescales in response to continuum variations, pointing to the origin of the Fe II emission in photoionized gas in the BLR. \cite{2012ApJS..202...10S} found that in a NLSy1 Ark 564, there is a a slightly better correlation of optical Fe II with the continuum at 5100 \AA\, than in the hydrogen Balmer lines, whereas in the case of another NLSy 1 galaxy NGC 4051, the variability of the optical Fe II emission also follows the continuum variability \citep{2005A&A...436..417W}. However, in other two cases, NGC 5548 \citep{2005ApJ...625..688V} and Ark 120 \citep{2008ApJ...673...69K}, weak correlation is seen, which might be due to poor cadence of the monitoring data, despite the large length of monitoring campaigns. Our findings may support the assumption that the central continuum emission is governing the production of Fe II lines \citep{2022AN....34310112G}. Based on the strong correlations between the Balmer lines and continuum luminosities, assuming that the continuum luminosity at 5100 \AA\, is a good tracer of ionization continuum, it is reasonable to assume that the Ly$\alpha$ line could correlate with the continuum luminosity at 5100 \AA. Therefore, we cannot rule out the possibility that continuum and L$\alpha$-pumping may be responsible for the excitation of Fe II upper levels (considering that this line is broad enough), which {then} populate the upper levels of transitions leading to optical Fe II. As already stated, some previous work emphasized the importance of Ly$\alpha$ fluoresence in Fe II production in AGN \citep[e.g.,][]{1987MNRAS.229P...1P,1998ApJ...499L.139S,2003ApJS..145...15S,2021ApJ...907...12S}. Some authors claimed that in NLSy1 other mechanisms, such as collisional excitation could be contributing as well to the Fe II production \citep[e.g.,][]{2000NewAR..44..531C}. However, here we see no difference in pop A and xA samples (which could be considered as representatives of NLSy1 objects) but even stronger dependence on the continuum luminosity. This is supported with strong correlations between the total Fe II emission with respect to the H$\alpha$ and H$\beta$ luminosities in both populations (Fig. \ref{fig:fe_lines}). Photoionization by the accretion disc {continuum} is also indicated by the correlation of Fe II emission with the accretion rate. Figure \ref{fig:fe_edd} shows total Fe II emission in pop A and pop B sample (left), and for xA sample (right) with respect to the Eddington ratio $L_{\rm Edd}/L_{\rm bol}$. {A} clear division between pop A and pop B objects is seen, {where} pop A {objects have} stronger Eddington ratio. {A} strong correlation of Fe II emission in the optical band with $L_{\rm Edd}/L_{\rm bol}$ is seen, which has been reported before \citep{2011ApJ...736...86D}.  Since the heating of the BLR plasma is dominantly through photoionization \citep{2006agna.book.....O}, this implies that the rate of collisions could be directly {proportional} to the input continuum ionizations.
Therefore,  with our findings we cannot rule out the effects of photoionization on the Fe II emission, influencing through all excitation mechanisms, i.e., collisional excitation, continuum and Ly$\alpha$ fluorescence. We believe that simultaneous observations of Fe II emission in AGN from UV to NIR are needed to understand the connection between these different physical processes and different Fe II emission.

Iron emission measured in three different wavelength bands, i.e. Fe II blue, Fe II green, and Fe II red is present in all objects, with Fe II (red) being somewhat weaker (Fig. \ref{fig:lumfe}, right panels). Pop B sample also contains objects with {the} strongest Fe II emission (red circled, Fig. \ref{fig:fe}, upper panels). This is not typically assumed, as strong Fe II emission is usually attributed to pop A (and NLSy1 as their subset). This could be the {result} of the fact that in these objects Fe II lines are broader and blended with underlying continuum or hidden in broad H$\alpha$ wings, and thus difficult to extract (see examples in Fig. \ref{fig:fit_examples}). This is why simultaneous multi-component fitting of AGN spectra, which is the approach implemented in \texttt{fantasy} code, may be important {in} studying the Fe II emission.

The most relevant finding of this analysis is the strong {correlation} between luminosity of Fe II in three bands (blue, green, red) for pop A and pop B samples (Fig. \ref{fig:fe}, upper panels). The strongest correlation is seen between the blue and green Fe II band (r=0.94 in pop A and pop B) which is expected since these bands are populated with most Fe II transitions (Figs. \ref{fig:grotrian} and \ref{fig:temp_dif}). These two bands surrounding H$\beta$ line are the ones typically measured and most Fe II templates are tackling these Fe II emission \citep{1992ApJS...80..109B, 2004A&A...417..515V, 2006ApJ...650...57T, 2022ApJS..258...38P}. However, the Fe II emission in the red band, redward from H$\alpha$ line, is also present and correlates with the iron emission present in the vicinity of H$\beta$ line (Fig. \ref{fig:fe}, left and right panels). When only xA sample is considered (Fig. \ref{fig:fe}, bottom panels), the correlations between Fe II bands are even stronger. 

Therefore, the red Fe II emission is contaminating the blue wing of H$\alpha$, whereas the red wing is significantly less affected (Fig. \ref{fig:fit_examples}). These iron blends are somewhat weaker and maybe hidden in the underlying continuum emission and broad H$\alpha$ wings in pop B objects (with broader emission lines), especially in low-quality spectra (i.e., poor S/N and spectral resolution). Some authors have been attributing this emission to broad line wings of H$\alpha$ line, for which {an} additional very broad line component had to be {introduced} \citep[as e.g., in][]{2017MNRAS.472.4051C}. This may significantly influence the measurements of H$\alpha$ flux and width, and their consequent application. For example, presence of Fe II emission could be responsible for the scatter of $M_{\rm BH}$ when measured from H$\alpha$ line \citep{Greene05, DallaBonta20}.

\begin{figure*}
    \centering
    \includegraphics[width=\textwidth]{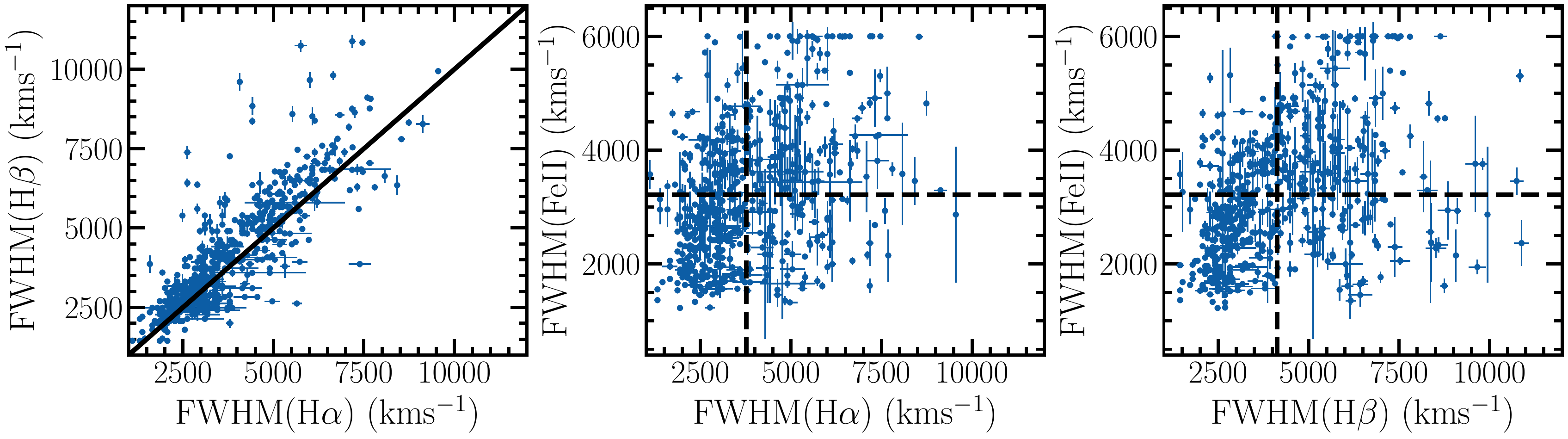}
    \caption{Width of the broad H$\beta$ plotted against the width of broad H$\alpha$ line (left) for the total SDSS sample, as well as the width of Fe II emission versus H$\alpha$ (middle) and H$\beta$ (right) broad line width. Line widths are in units of km s$^{-1}$. Linear best-fitting line is displayed in H$\beta$ vs. H$\alpha$ plot, whereas on the two plots dashed {lines} denote the average values for the total sample. }
    \label{fig:fwhm}
\end{figure*}

Finally, we check the kinematics of Fe II lines with respect to the broad H$\alpha$ and H$\beta$ lines through the analysis of the line FWHM. Figure \ref{fig:fwhm} presents the width of Fe II emission versus H$\alpha$ (middle) and H$\beta$ (right) broad line width for the total SDSS sample. The average width of Fe II lines in total sample is $\sim$3300 km s$^{-1}$, whereas H$\alpha$ FWHM is $\sim$3780 km s$^{-1}$ and H$\beta$ is $\sim$4120 km s$^{-1}$. The two broad components used to fit the H$\alpha$ line have the widths of 6640 km s$^{-1}$ (6300 km s$^{-1}$ for H$\beta$) and 2470 km s$^{-1}$ (2450 km s$^{-1}$ for H$\beta$). This is in agreement with {the} previous assumption that Fe II emission is originating in so called intermediate line region \citep{kov10,2011ApJ...736...86D}.

\subsection{Fittings of type 1 AGN spectra with \texttt{fantasy} code}

Using the \texttt{fantasy} code, we successfully modeled the AGN optical spectra ($\lambda\lambda$ 4000-7000 \AA) by fitting all emission components simultaneously: the underlying broken power-law continuum, broad and narrow emission lines, and the Fe II model (Fig. \ref{fig:fit_examples}). The observed strong correlations between the line and continuum luminosities, and between the broad line luminosities indicate good performance of the automatic spectral fitting with the \texttt{fantasy} code. This is also confirmed by the strong correlation between the FWHMs of H$\alpha$ and H$\beta$ line (Fig. \ref{fig:fwhm}, left panel).  We note that in few cases ($<3$\%) the width of Fe II lines was pegged to the upper limit set by the code (Fig. \ref{fig:fwhm}, middle and right panels). These objects are pop B with very strong Fe II emission in all three bands. We visually inspected these fits, and found that the broad H$\alpha$ and H$\beta$, and Fe II bands were well fitted, from which we concluded that line fluxes have been correctly extracted. These could generally be removed from the analysis, or fitted with models that have lower constraints on the limits, but, for consistency, we decided to include them, as they are not influencing the presented results.

In addition, we have shown that especially when investigating the H$\alpha$ line, the use of the \texttt{fantasy} code can be important to detect the Fe II emission hidden in the H$\alpha$ wings and to carefully measure its spectral parameters. Overall, our findings show that the \texttt{fantasy} code is well suited for modeling SDSS type 1 AGN spectra. Finally, we have demonstrated that even in case of strong iron emitters, such as I Zw 1, the fittings using all emission components reproduced well the observed spectrum, especially in the vicinity of H$\alpha$ line (Fig. \ref{fig:fitZw}).

In addition to other well specialized program packages for AGN spectral fittings, namely pyQSOFit \citep{{2018ascl.soft09008G, 2019MNRAS.482.3288G}} and QSfit \citep{2017MNRAS.472.4051C}, which are widely used in analysis of SDSS data \citep[see e.g.,][]{2011ApJS..194...45S,2019ApJS..241...34S,2020ApJS..249...17R}, the \texttt{fantasy} code seems to be also well suited to decompose the continuum and various emission features in AGN spectra. Its useful features are that it is user-friendly, contains necessary procedures for pre-processing and preparing the spectra for spectral decomposition, and the approach to simultaneously fit the emission lines and underlying continuum. In addition,
{the predefined lists of possible emission lines in AGN (e.g. strong narrow lines, see Table \ref{tab:lines} in the Appendix)}, models with emission lines with fixed parameters (width, shift, line ratios), and model of Fe II emission, make \texttt{fantasy} a unique tool.  However, apart from {a} numerical estimate of uncertainties from the fittings using the Monte Carlo approach, \texttt{fantasy} lacks a more serious treatment of uncertainties, which should be addressed in the future.

Nevertheless, the features listed above (e.g., simultaneous fitting of different components, predefined line lists, flexibility to model a wide variety of spectra) make this code well suited for modeling optical AGN spectra. Of particular interest are the modeling of optical spectra in transient events, such as the strong iron TDEs, which was shown to be successful by \cite{2023A&A...669A.140P}.

\section{Summary}

We present here the study of the physical properties of broad line emitting regions in type 1 AGN, with special attention to Fe II emission in the wavelength range of the H$\alpha$ line.
We use a sample of 655 objects from the current SDSS DR17, selected to cover the wavelength range 4000-7000 \AA\, (i.e., containing both H$\alpha$ and H$\beta$ lines). Our goal is to analyze only high-quality spectra (S/N ratio $>$35) so that we can reliably measure Fe II emission around H$\beta$ but also near the H$\alpha$ line, where it is typically blended with underlying continuum and broad H$\alpha$ wings, particularly in low S/N spectra. We present an updated approach to multicomponent fitting of AGN spectra using the \texttt{Python} code \texttt{fantasy}. We present an extended model of Fe II emission based only on the atomic data, covering the wavelength range near the H$\alpha$ line, which has been poorly studied in the past. We perform spectral fitting using the code \texttt{fantasy}, which allowed us to measure the spectral parameters of the broad H$\gamma$, H$\beta$ and H$\alpha$ lines from the pure AGN spectra, as well as the iron emission in three bands: Fe II blue (4340--4680 \AA), Fe II green (5100--5600 \AA), and Fe II red (6100--6650 \AA). 

Our main conclusions can be summarized as follows:
\begin{enumerate}
    \item The Fe II emission, if present in the vicinity of H$\beta$ line, is also detected redward from H$\alpha$ line, at a comparable strength. This red Fe II emission contaminates the broad H$\alpha$ line red-wing, which can have an impact on the measured H$\alpha$ flux and width. This can be particularly important for pop B objects (with broader emission lines), since iron blends near the H$\alpha$ line are hidden in the underlying continuum and broad H$\alpha$ wings; 
    \item The production of Fe II emission is strongly correlated with Eddington luminosity, and appears to be controlled by the same mechanism as the hydrogen Balmer lines, as shown by strong correlations with continuum and line luminosities. This implies that photionization is governing the Fe II production, however the exact mechanism (e.g., collisional excitation, continuum or Ly$\alpha$ fluorescence) is not constrained through this analysis;
    \item Simultaneous multicomponent fitting of complex AGN spectra is a necessary approach for broad line parameter extraction, especially for reliable measurement of H$\alpha$ spectral parameters (width and flux). The open-source code \texttt{fantasy} tested in this work appears to be well suited for modeling the spectrum of type 1 AGN and may prove useful in future studies of a large number of AGN spectra.
\end{enumerate}

To date, most Fe II templates have focused on the H$\beta$ line, which is one of the best studied AGN emission lines. However, with current and future high-precision instruments that will focus more on the NIR spectrum, observation of the H$\alpha$ line in distant quasars will be more present, and more Fe II templates and models in this wavelength range will be needed.

\begin{acknowledgments}
Authors would like to thank the anonymous referee whose comments and suggestions helped to improve this manuscript. D.I. and L.\v C.P. acknowledge funding provided by the University of Belgrade - Faculty of Mathematics (the contract \textnumero451-03-47/2023-01/200104) and Astronomical Observatory Belgrade (the contract \textnumero451-03-47/2023-01/200002) through the grants by the Ministry of Science, Technological Development and Innovation of the Republic of Serbia. D.I. acknowledges the support of the Alexander von Humboldt Foundation.

This research uses data from the SDSS Data Release 17 \citep{dr17}. Funding for the Sloan Digital Sky Survey IV has been provided by the 
Alfred P. Sloan Foundation, the U.S. Department of Energy Office of 
Science, and the Participating Institutions. SDSS-IV acknowledges support and resources from the Center for High Performance Computing  at the University of Utah. The SDSS website is \url{www.sdss4.org}.

This research has made use of the NASA/IPAC Extragalactic Database (NED), which is operated by the Jet Propulsion Laboratory, California Institute of Technology, under contract with the National Aeronautics and Space Administration.
\end{acknowledgments}

\vspace{10mm}
\facilities{SDSS \citep{dr17}, NED, NIST \citep{NIST_ASD}, SUPERAST \citep[][]{2022PASRB..22..231K}.}

\software{astropy \citep{2013A&A...558A..33A,2018AJ....156..123A},  
          sherpa \citep{doug_burke_2022_7186379}, fantasy \citep{ilic20,rakic22}, 
          sfdmap \citep{schlegel98},
          PyAstronomy \citep{pya},
          spectres \citep{2017arXiv170505165C}.
          }

\appendix

\section{\texttt{fantasy}~Predefined line lists}


We provide all predefined lists of standard AGN emission lines
    {present} in the \texttt{fantasy} code. These are: Hydrogen
    lines (Balmer and Paschen series), Helium lines (both He I and He
    II), strongest narrow emission lines ([O III], [N II]), other AGN
    narrow lines (e.g., [S II], [O I]), other AGN broad lines (Ca I,
    O I), coronal lines (e.g. [Fe X], [Ar V]), and forbidden Fe II
    lines (Table \ref{tab:lines}).  The presented list is very
    extensive, and most of the lines are not detected in AGN
    \citep[see e.g., mean quasar spectrum from][]{vanden_berk06}.
    However, we believe that such a comprehensive single list of
    possible emission lines is useful.

When {constructing} these {line lists} (Table
    \ref{tab:lines}) we acknowledge the usage of a collection of
    detected emission lines in galaxies compiled by S. Drew
    Chojnowski\footnote{\url{http://astronomy.nmsu.edu/drewski/
    tableofemissionlines.html}}, the NIST database \citep{NIST_ASD},
    {narrow} lines listed in \cite{2004A&A...417..515V} and
    \cite{2022ApJS..258...38P}, {some of the solar coronal lines
    \citep[][]{2018ApJ...852...52D},} as well as other papers cited
    in Section 3.

{For all listed lines, we provide the air wavelengths}.
Full data sets are also available in the machine readable format.

\begin{deluxetable}{lccc}
\tabletypesize{\scriptsize}
\tablewidth{0pt} 
\tablecaption{Predefined {lists} of emission lines used in \texttt{fantasy} code.\label{tab:lines}}
\tablehead{
\colhead{Line} & \colhead{Wavelength {(air)}} & \colhead{Type} & \colhead{Group} \\
 &  \colhead{\AA}
} 
\startdata 
H$\epsilon$ & 3970.08 & narrow, broad & hydrogen\\
H$\delta$ & 4101.74 & narrow, broad & hydrogen\\
H$\gamma$ & 4340.47 & narrow, broad & hydrogen\\
H$\beta$ & 4861.33 & narrow, broad & hydrogen\\
H$\alpha$ & 6562.82 & narrow, broad & hydrogen\\
Pa14 & 8598.39 & narrow, broad & hydrogen\\
Pa13 & 8665.02 & narrow, broad & hydrogen\\
Pa12 & 8750.47 & narrow, broad & hydrogen\\
Pa11 & 8862.78 & narrow, broad & hydrogen\\
Pa10 & 9014.91 & narrow, broad & hydrogen\\
Pa9 & 9229.01 & narrow, broad  & hydrogen\\
Pa$\epsilon$  & 9545.97 & narrow, broad & hydrogen \\
Pa$\delta$  & 10049.37 & narrow, broad & hydrogen\\
Pa$\gamma$ & 10938.09 & narrow, broad & hydrogen\\
\hline
\ldots & \ldots & \ldots & \ldots \\
\enddata
\tablecomments{Columns give line wavelength in air (in \AA), {a}
    type of line in terms of line width expected in AGN spectra, and
    {a} group assigned within {a} predefined list in
    \texttt{fantasy} code. Table \ref{tab:lines} is published in its
    entirety in the machine-readable format. A portion is shown here
    for guidance regarding its form and content.}
\end{deluxetable}

\typeout{}\clearpage
\bibliography{new.ms}{}
\bibliographystyle{aasjournal}

\end{document}